\documentclass[12pt]{article}
\pdfoutput=1

\usepackage[utf8]{inputenc}

\usepackage{draft}
\usepackage{putex}

\usepackage[all]{xy}

\usepackage{stackrel,amssymb}

\usepackage{graphicx}
\usepackage{caption}
\usepackage{amsmath,bm}
\usepackage{array}
\usepackage{subcaption}
\usepackage{epstopdf}
\usepackage{enumerate}
\usepackage{cite}
\usepackage{tensor}
\usepackage{slashed}
\usepackage[utf8]{inputenc}
\usepackage{rotating}
\usepackage{bigfoot}
\usepackage[
colorlinks=true,
linkcolor=blue,
urlcolor=blue,
filecolor=black,
citecolor=red,
]{hyperref}

\usepackage{tikz-cd}

\def\XXint#1#2#3{{\setbox0=\hbox{$#1{#2#3}{\int}$ }
		\vcenter{\hbox{$#2#3$ }}\kern-.6\wd0}}

\usepackage{adjustbox}
\usepackage{multirow}
\usepackage{tikz-cd}

\numberwithin{equation}{section}

\newcommand{\CC}{\Gamma}
\newcommand{\cI}{{\cal I}}

\newtheorem{theorem}{Theorem}
\newtheorem{conj}{Conjecture}

\def\<{\langle}
\def\>{\rangle}

\def\pa{\partial}

\def\ep{\epsilon}
\def\sD{\slashed{D}}

\newcommand*\DA{\mathop{}\!\mathbin\Box}

\newcommand{\leftrarrows}{\mathrel{\raise.75ex\hbox{\oalign{%
				$\scriptstyle\leftarrow$\cr
				\vrule width0pt height.5ex$\hfil\scriptstyle\relbar$\cr}}}}
\newcommand{\lrightarrows}{\mathrel{\raise.75ex\hbox{\oalign{%
				$\scriptstyle\relbar$\hfil\cr
				$\scriptstyle\vrule width0pt height.5ex\smash\rightarrow$\cr}}}}
\newcommand{\Rrelbar}{\mathrel{\raise.75ex\hbox{\oalign{%
				$\scriptstyle\relbar$\cr
				\vrule width0pt height.5ex$\scriptstyle\relbar$}}}}

\makeatletter
\def\leftrightarrowsfill@{\arrowfill@\leftrarrows\Rrelbar\lrightarrows}
\newcommand{\xleftrightarrows}[2][]{\ext@arrow 3399\leftrightarrowsfill@{#1}{#2}}
\makeatother

\begin{document}
	
	\preprint{}
	
	\institution{CMSA}{Center of Mathematical Sciences and Applications, Harvard University, Cambridge, MA 02138, USA}
	\institution{HU}{Jefferson Physical Laboratory, Harvard University,
		Cambridge, MA 02138, USA}

	\title{
		Defect $a$-Theorem and $a$-Maximization
	}

	\authors{Yifan Wang\worksat{\CMSA,\HU}}

	\abstract{
		Conformal defects describe the universal behaviors of a conformal field theory (CFT) in the presence of a boundary or more general impurities. The coupled critical system is characterized by new conformal anomalies which are analogous to, and generalize those of standalone CFTs. Here we study the conformal $a$- and $c$-anomalies of four dimensional defects in CFTs of general spacetime dimensions greater than four. We prove that under unitary defect renormalization group (RG) flows, the  defect $a$-anomaly must decrease, thus establishing the defect $a$-theorem. 
		For conformal defects preserving minimal supersymmetry, the full defect symmetry contains a distinguished $U(1)_R$ subgroup. 
		We derive the anomaly multiplet relations that express the defect $a$- and $c$-anomalies in terms of the defect (mixed) 't Hooft anomalies for this $U(1)_R$ symmetry. 
		Once the $U(1)_R$ symmetry is identified using the defect $a$-maximization principle which we prove,  this enables a non-perturbative pathway to the conformal anomalies of strongly coupled defects.
		We illustrate our methods by discussing a number of examples including boundaries in five dimensions and codimension-two defects in six dimensions. We also comment on chiral algebra sectors of defect operator algebras and potential conformal collider bounds on defect anomalies.

	}
	\date{}

	\maketitle
	
	\tableofcontents

	\pagebreak
	
	\section{Introduction and Summary}
	The \textit{theory space} of quantum field theories (QFT) is immensely rich and diverse. It encompasses a wide range of quantum dynamics, including renormalization group (RG) flows, phase transitions and critical phenomena, that take place in a large and elaborate zoo of quantum systems.
	An enduring challenge is to identify and understand non-perturbative structures in the geometry of the theory space.

	A particularly well-posed and important problem is the monotonicity of RG flows. This is intuitively clear since the RG procedure involves coarse-graining over short-distance physics and therefore leads to a reduction in the effective degrees of freedom. 
	In spacetime dimension $d=2$, this intuitive picture was rigorously justified by the celebrated $c$-theorem \cite{Zamolodchikov:1986gt}, where a height function, known as the $c$-function, was constructed over the theory space, which decreases monotonically under RG flows. 
	A particular feature of the $2d$ $c$-function is that it coincides with the conformal anomalies at the critical fixed points of the RG flows, described by conformal field theories (CFT). Such conformal anomalies are expected to ``count'' the degrees of freedom in the CFTs, thus it is natural to study conformal anomalies in higher dimensions in an effort to extend the $2d$ $c$-theorem. 
	
	The conformal anomalies of CFTs are present for even $d$ and important physical observables that govern the CFT dynamics (e.g. through correlation functions of the stress-tensor). They are also known as the trace anomalies, since the symmetric, conserved, and traceless stress-tensor $T_{\m\n}$ in a CFT $\cT$ can develop an anomalous nonzero trace when the theory is placed on a spacetime manifold with nontrivial metric $(\cM,g)$ \cite{Deser:1993yx},\footnote{We only include the contributions that cannot be removed by adjusting local counter-terms.}
	\ie
	\cA^{\rm W}_\cT\equiv \la T^\m_\m \ra= {1\over (4\pi)^{d\over2}}\left(-(-1)^{d\over 2} a E_d +\sum_i c_i W_i\right)\,,
	\label{trgen}
	\fe
	that solves the Wess-Zumino consistency conditions \cite{Wess:1971yu,Osborn:1991gm}.
	Equivalently, the conformal anomalies  contribute to the anomalous variation of the path integral, under the Weyl transformation $g_{\m\n}\to e^{2\sigma(x)}g_{\m\n}$,
	\ie
	\D_\sigma \log Z[g]= i\int_\cM d^d x \sqrt{|g|} \,\sigma  \cA^{\rm W}_\cT\,.
	\label{weylvarZ}
	\fe
	Here $E_d$ is the Euler density in $d$ dimensions, which is related to the Euler characteristic by $\chi(\cM)={1\over 2(4\pi)^{d/2}}\int_\cM E_d$ where $\chi(S^d)=2$ for an even dimensional sphere, and $W_i$ are Weyl invariants of the Riemann curvature (that are not total derivatives) of scaling dimension $d$.\footnote{There are no such Weyl invariants for $d=2$, one for $d=4$ and three for $d=6$.}
	The $c$-anomaly in $d=2$ is proportional to the $a$-coefficient in \eqref{trgen} and \eqref{weylvarZ}. It is thus natural to expect the $a$-anomaly to play the role of the 2d $c$-anomaly for general even spacetime dimensions and the monotonicity of RG flows to be governed by an $a$-theorem  \cite{Cardy:1988cwa}.
	
	Decades after the work of \cite{Zamolodchikov:1986gt}, the $a$-theorem was finally proven in $d=4$ \cite{Komargodski:2011vj,Komargodski:2011xv,Casini:2017vbe}. A key insight of \cite{Komargodski:2011vj,Komargodski:2011xv} is a conformal version of the conventional spurion analysis for global symmetries. Upon introducing a background dilaton field suitably coupled to the theory of interest, one can restore conformal symmetry along the RG flow. Anomaly matching then suggests that the difference between the UV and IR conformal anomalies must be reproduced by the Weyl transformation of the spurious dilaton. The $d=4$ $a$-theorem then follows from unitarity constraints on the dilaton effective action. Subsequent efforts to generalize the argument of  \cite{Komargodski:2011vj,Komargodski:2011xv} to higher dimensions were  made in $d=6$  \cite{Elvang:2012st,Elvang:2012yc,Cordova:2015vwa,Cordova:2015fha}. Here we focus on a different extension, namely the $a$-theorem for conformal defects of dimension $p=4$ in $d>4$ dimensional CFTs. Closely related is the proof of $b$-theorem for surface defects (i.e. $p=2$) in \cite{Jensen:2015swa} (see also \cite{Wang:2020xkc}).
	
	Defects are an integral part of modern understanding of QFT. Familiar examples include boundary conditions and quantum impurities (e.g. Wilson line of a probe particle in a gauge theory). More generally defects define extended operators over submanifolds of the spacetime, which enrich the  algebra of local operators. On the one hand, they participate actively in the bulk field theory dynamics, providing elegant formulations of (generalized) symmetries and behaving as non-local order parameters for bulk phase transitions \cite{Gaiotto:2014kfa}.
	On the other hand, they give rise to new phase transitions and critical phenomena localized on the $p$-dimensional defect worldvolume. The incorporation of defects is clearly essential for a proper understanding of the theory space of QFTs. 
	
	Similar to how CFTs correspond to  bulk universality classes in the theory space, conformal defects  describe universality classes of defect RG flows. The coupled bulk-defect critical phase is also commonly referred to as the defect CFT (DCFT) (see \cite{Andrei:2018die} for a recent review). A $p$-dimensional conformal defect $\cD$ in a $d$-dimensional CFT shares many kinematic features of a standalone $p$-dimensional CFT, for the obvious reason that they are both invariant under the $SO(p,2)$ conformal symmetry. However a crucial difference of a defect $\cD$ from a standalone CFT in $p$-dimensions is the generic absence a locally conserved $p$-dimensional stress-tensor. Instead, the defect symmetry is inherited from the ambient CFT. From now on we focus on the case $p=4$, namely the conformal defect $\cD$ corresponds to a boundary (interface) in a $5d$ CFT or a higher codimension defect in $d>5$ CFTs.
	
	The conformal anomalies of usual CFTs \eqref{trgen} generalize to these DCFTs in a straightforward fashion \cite{Graham:1999pm,Henningson:1999iw,Solodukhin:2015eca}. The bulk stress-tensor receives, in addition to its usual trace anomaly $\cA_\cT^{\rm W}$ in the absence of defects \eqref{trgen}, anomalous trace contributions $\cA^{\rm W}_\cD$ localized on the defect worldvolume $\Sigma \subset \cM$,
	\ie
	\la T^\m_\m (x) \ra_\cD = \cA^{\rm W}_\cT +  \D(\Sigma)  \cA^{\rm W}_\cD\,,
	\quad 
	\cA^{\rm W}_\cD = {1\over (4\pi)^2}\left (-a E_4 +c W \right) + I_{\rm ext}(C^{(d)}, K)\,.
	\label{dta}
	\fe
	As before, the same anomalies are captured by the anomalous variation of the DCFT partition after coupling to background geometry,
	\ie
	\D_\sigma \log Z_\cD [g,X]=i\int_\cM d^d x \sqrt{|g|}\,\sigma \cA^{\rm W}_\cT+i \int_\Sigma d^4 z \sqrt{|h|} \,\sigma \cA^{\rm W}_\cD\,,
	\label{dva}
	\fe 
	under a Weyl transformation of the ambient metric $g_{\m\n}(x) \to e^{2\sigma(x)} g_{\m\n}(x)$.
	Here $z^a$ are coordinates on the submanifold $\Sigma \subset \cM$ which is specified by the embedding functions $X^\m(z^a)$.\footnote{We assume that the normal bundle of $\Sigma$ is topologically trivial, which is obviously the case for the conformal defect in flat space.} The induced metric on $\Sigma$ is $h_{ab}=\pa_a X^\m \pa_b X^\n g_{\m\n}$ which transforms as $h_{ab}(z) \to e^{2\sigma(X(z))}h_{ab}(z)$\, under the Weyl transformation.

	A few remarks are in order. In \eqref{dta} and \eqref{dva},  $E_4$ and $W$ are the intrinsic Euler density and quadratic Weyl invariant of $\Sigma$ which also appear in the trace anomaly of a standalone 4d CFT, and we refer to them as the \textit{intrinsic conformal anomalies} of the defect $\cD$. They are given explicitly by the following combinations of Riemann curvatures on $\Sigma$,
	\ie
	E_4=& R_{abcd}R^{abcd}-4R_{ab}R^{ab}+R^2 = {1\over 4} \ep^{abcd} \ep_{efgh} R^{ef}{}_{ab}R^{gh}{}_{cd}\,,
	\\
	W=&R_{abcd}R^{abcd}-2R_{ab}R^{ab}+{1\over 3}R^2  =C_{abcd}C^{abcd}\,.
	\fe
	We will continue to use $a$ and $c$ to denote the corresponding conformal anomaly coefficients of the DCFT. Note that these anomalies are present and universal regardless of bulk spacetime dimension (e.g. $d$ can be odd). The full structure of conformal anomalies for the DCFT is however much richer then that in a standalone 4d CFT, with extra contributions corresponding to \textit{extrinsic conformal anomalies} of the defect $\cD$, given by the last term $I_{\rm ext}(C^{(d)},K)$ in \eqref{dta}.\footnote{See \cite{Speranza:2019hkr} for general discussions of geometric invariants for submanifolds and their relations through Gauss-Codazzi type identities.} It contains additional diffeomorphism invariants that depend on the embedding $\Sigma\subset\cM$ subject to the Wess-Zumino consistency condition \cite{Wess:1971yu,Osborn:1991gm}, and in particular includes Weyl invariants constructed from the (traceless) extrinsic curvature $K^\m_{ab}$ and bulk Weyl curvatures $C^{(d)}_{\m\n\rho\sigma}$ pulled back to $\Sigma$. The independent extrinsic conformal anomalies for a $p=4$-dimensional defect have not been completely classified (see \cite{Henningson:1999xi} for a partial list in the $d=5$ case\footnote{We thank Sergey Solodukhin for correspondence on this point.}) but their explicit forms will not be important for this work.\footnote{Related works on conformal invariants of $p=4$ submanifolds include the Willmore energy in \cite{Zhang,Graham:2017bew} which generalizes the extrinsic Graham-Witten conformal anomaly for surface defects in \cite{Graham:1999pm}. See also  \cite{Mondino} for  a partial classification of submanifold conformal invariants for general $p$ and $d$ in \cite{Mondino}.
	} 
	
	A main question of interest here is whether the  defect $a$-anomaly in \eqref{dta} decreases monotonically under defect RG flows. In Section~\ref{sec:dathm}, we find an affirmative answer by extending the work of \cite{Komargodski:2011vj,Komargodski:2011xv} and invoking Lorentzian unitarity constraints on the defect dilaton effective action, thus establishing the defect $a$-theorem. We also describe explicitly the defect dilaton effective action for the simple RG flow between conformal Neumann and Dirichlet boundary conditions in the $d=5$ free scalar theory. 
	
	If we know the defect $a$-anomaly of a UV DCFT or an IR DCFT at either ends of a defect RG flow,  the defect $a$-theorem produces non-perturbative constraints on the RG trajectory. However such conformal anomalies are notoriously difficult to access, already in standalone CFTs. For $p=4$ DCFTs, our knowledge is even more limited: the defect $a$-anomalies were only known for the free scalar theory \cite{Rodriguez-Gomez:2017aca,Rodriguez-Gomez:2017kxf,Nishioka:2021uef}\footnote{We thank Yoshiki Sato for correspondence on this point.} and for the free fermion in $d=5$ \cite{Rodriguez-Gomez:2017aca}, and so far no results for the defect $c$-anomalies have been obtained. We will address these issues in this work.
	For $d=4$ CFTs with $\cN=1$ supersymmetry (SUSY), an elegant non-perturbative method was developed to solve for the $a$- and $c$-anomalies, known as $a$-maximization \cite{Intriligator:2003jj}. The $\cN=1$ SUSY relates in a simple way the conformal $a$- and $c$-anomalies to the 't Hooft anomalies involving the $U(1)_R$ symmetry of the CFT \cite{Anselmi:1997am}. The 't Hooft anomalies are much easier to compute thanks to their robustness under deformations (which can break conformal but preserve Poincar\'e and $U(1)_R$ symmetries and possibly other flavor symmetries). The only subtlety is to identify the $U(1)_R$ symmetry, which can mix with other $U(1)$ flavor symmetries. This is accomplished by the $a$-maximization principle \cite{Intriligator:2003jj,Kutasov:2003iy,Kutasov:2003ux}.\footnote{In \cite{Barnes:2005bm}, another way to identify the superconformal $U(1)_R$ symmetry was introduced, known as $\tau_{RR}$-minimization, by minimizing the R-current two-point function. The $\tau_{RR}$-minimization and $a$-maximization are equivalent at the fixed point, but in practice the latter is more powerful  since the 't Hooft anomalies are well-defined away from the CFT.} 
	
	For the $p=4$ conformal defects preserving the minimal $\cN=1$ SUSY, in Section~\ref{sec:daamax} we show that the same relations between conformal anomalies and 't Hooft anomalies in 4d SCFTs \cite{Anselmi:1997am} continue to hold for the DCFTs. The $U(1)_R$ superconformal R-symmetry in the DCFT descends from the (larger) R-symmetry and transverse rotation symmetry of the ambient SCFT. The relevant 't Hooft anomalies amount to a modification of the Ward identity of the $d$-dimensional $U(1)_R$ current $J$ on the defect worldvolume $\Sigma$,
	\ie
	\la \nabla_\m J^\m \ra_\cD \supset {\D(\Sigma)\over 2(2\pi)^2} \star_\Sigma \left(-{1\over 3} k_{ RRR} F\wedge F+{1\over 24} k_{R} \tr \cR\wedge \cR \right) \,,
	\fe
	where $F=dA$ is the background $U(1)_R$ field strength and $\cR$ is the Riemann curvature two-form. 
	The anomaly coefficients above are simply related to the defect conformal anomalies by
	\ie
	a={9k_{RRR}-3k_R\over 32},\quad c={9k_{RRR}-5k_R\over 32}\,,
	\fe
	as in the standalone SCFT \cite{Anselmi:1997am}. Furthermore in the presence of 4d conserved currents on the defect, we prove a defect version of the $a$-maximization principle \cite{Intriligator:2003jj} (see also \cite{Barnes:2005bm}) that identifies the superconformal $U(1)_R$ symmetry as a linear combination that may involve these 4d currents. As a by-product, this defect $a$-maximization principle also leads to an alternative non-perturbative proof of the defect $a$-theorem for certain supersymmetric RG flows, extending the results of \cite{Kutasov:2003iy,Barnes:2004jj}.
	
	In Section~\ref{sec:examples}, to illustrate our methods, we apply the defect $a$-maximization principle to selected examples of 4d $\cN=1$ DCFTs in 5d and 6d SCFTs, and determine their defect conformal anomalies. For free theories, our results are consistent with the previous answers on defect $a$-anomalies \cite{Rodriguez-Gomez:2017aca,Rodriguez-Gomez:2017kxf,Nishioka:2021uef} obtained from the heat kernel method \cite{Vassilevich:2003xt} and produce new constraints on their defect $c$-anomalies. For interacting theories in 5d, we discuss boundary conditions of 5d SCFTs, including a boundary version of the familiar 4d $\cN=1$ SQCD and its boundary conformal anomalies. In 6d, we give a reinterpretation of the known results of the ``conformal anomalies of punctures''  in (generalized) class S constructions \cite{Gaiotto:2009we,Gaiotto:2009gz,Gaiotto:2009hg} as the conformal anomalies of codimension-two defects as defined in \eqref{dta}.
	We end with a brief discussion of future directions in Section~\ref{sec:discussion}.
	
	\section{Defect $a$-Theorem}
	\label{sec:dathm}
	
	In this section, we will prove the following defect $a$-theorem that constrains RG flows between conformal defects of dimension $p=4$.
	\begin{theorem}[Defect $a$-Theorem]
		For a  unitary defect RG flow between two unitary conformal defects $\cD_{\rm UV}$ and $\cD_{\rm IR}$ of dimension $p=4$, the corresponding defect $a$-anomalies satisfy
		\ie
		a({\cD_{\rm UV}}) > a({\cD_{\rm IR}})\,. 
		\fe
		\label{thm:athm}
	\end{theorem}
	In analogy to the different versions of $a$-theorems for 4d QFTs, the above would be the \textit{weak} version of the defect $a$-theorem. A \textit{stronger} version of the $a$-theorem requires an $a$-function defined on the entire theory space which has the desired monotonicity properties and coincides with the conformal $a$-anomalies at the fixed points. The \textit{strongest} $a$-theorem further demands the RG flows to be gradient flows with respect to the $a$-function.  Although the stronger (and strongest) $a$-theorem has not been completely proven, there is ample evidence for its validity (see e.g. \cite{Osborn:1989td,Jack:1990eb,Freedman:1998rd,Freedman:1999gp,Barnes:2004jj}). Similarly one can formulate a stronger version of the defect $a$-theorem below (analogously for the strongest version), but a proof is beyond the scope of this work. Closely related is the question whether scale invariance implies conformal invariance for $p=4$ defects. Substantial progress has been made in proving their equivalence in unitary interacting 4d QFTs \cite{Luty:2012ww,Fortin:2012hn,Nakayama:2013is, Bzowski:2014qja,Dymarsky:2013pqa,Dymarsky:2014zja,Yonekura:2014tha,Naseh:2016maw}, which should also have natural extensions to defects. 
	\begin{conj}
		Given a unitary CFT $\cT$ of dimension $d>4$, there exists a height function ($a$-function) ${\bm a}(\lambda_i)$ on the space of $p=4$ unitary Poincar\'e invariant defects in $\cT$, parametrized by couplings $\{\lambda_i\}$, such that under a defect RG flow parametrized by scale $\m$,
		\ie
		\mu {d\over d\mu} {\bm a}(\lambda_i)=\B_j(\lambda_i) {\pa\over \pa \lambda_j}{\bm a}(\lambda_i) \geq 0\,.
		\fe
		Moreover the inequality is saturated at the fixed points $\lambda_i=\hat\lambda_i$ with $\B_j(\hat\lambda_i )=0$, where the value of ${\bm a}$ coincides with the defect conformal $a$-anomaly of the fixed point DCFT   
		\ie
		{\bm a}(\hat\lambda_i)=a(\cD)\,.
		\fe
	\end{conj}
	We emphasize that the $a$-function here is subject to the \textit{local} condition, i.e. the $\mu$ dependence of ${\bm a}(\lambda_i)$ comes entirely from the running couplings $\lambda_i$.

	\subsection{Monotonicity theorem from dilaton effective action}
	Here we prove the defect $a$-theorem (Theorem~\ref{thm:athm}) by extending the method of \cite{Komargodski:2011vj,Komargodski:2011xv} (see also \cite{Luty:2012ww}). We will see that apart from a few novelties due to the difference between a defect and a standalone CFT (e.g. the extra extrinsic anomalies in \eqref{dta}), the arguments in \cite{Komargodski:2011vj,Komargodski:2011xv,Luty:2012ww} essentially carry through (which we explain to make it self-contained), and the theorem is established by a defect version of the optical theorem.

	We start by considering a defect RG flow from a UV conformal defect $\cD_{\rm UV}$ to an IR conformal defect $\cD_{\rm IR}$. The defect conformal symmetry is explicitly broken in the defect field theory (DFT) at an immediate scale but we can restore it by coupling to a non-dynamical dilaton field $\tau(z)$ on the defect worldvolume $\Sigma$ which transforms as $\tau \to \tau +\sigma$  under a local Weyl rescaling of the ambient metric $g_{\m\n} \to e^{2 \sigma} g_{\m\n}$.\footnote{As usual, one can think of conformal symmetry as the subgroup of ${\rm Diff}\times{\rm Weyl}$ that leaves the flat space metric invariant.}  
	
	Denoting the UV DCFT action abstractly by $S_{\cD_{\rm UV}}$, the DFT is generally described by a relevant deformation on the defect worldvolume $\Sigma$,
	\ie
	S_{\rm DFT}=S_{\cD_{\rm UV}}+\int_\Sigma d^4 z \sqrt{|h|} \,\sum_{\cO_{\rm UV} \in \cD_{\rm UV}}\lambda_{\cO_{\rm UV}} \cO_{\rm UV}(z)\,.
	\fe
	After coupling to  background dilaton, we have
	\ie
	S_{\rm DFT}[\tau]=S_{\cD_{\rm UV}}+\int_\Sigma d^4 z \sqrt{|h|} \sum_{\cO_{\rm UV}\in \cD_{\rm UV}} \Omega^{4-\Delta_{\cO_{\rm UV}}}\lambda_{\cO_{\rm UV}}  \cO_{\rm UV}(z)\,,
	\label{UVdS}
	\fe
	with
	\ie
	\Omega\equiv e^{-\tau}\,,
	\fe
	and the Weyl invariance becomes manifest. 
	To linearized order, the coupling takes the form
	\ie
	S_{\rm DFT}[\tau] \equiv S_{\rm DFT}+ \int_\Sigma d^4 z \sqrt{|h|} \,T(z) \tau(z) + \cO(\tau^2)\,,
	\fe
	where $T(z)$ is an operator coming from the trace of the bulk stress-tensor $T^\m_\m(x)=\D(\Sigma)T(z)$, which is nontrivial away from the defect fixed points.\footnote{Near the UV fixed point, $T(z)$ is dominated by the relevant (or marginally relevant) operator $\cO_{\rm UV}$ with the largest scaling dimension $\Delta_{\cO_{\rm UV}} \leq 4$.} 
	
	Having reinstated the Weyl symmetry with a compensator field $\tau$, the anomalous Weyl variation of the partition function $Z_{\rm DFT}[\tau]$ must be constant along the RG flow as a consequence of the Wess-Zumino consistency condition and thus completely determined by the conformal anomalies of the UV DCFT $\cD_{\rm UV}$. Consequently, near the IR end of the defect RG flow, the same Weyl variation must be reproduced by the effective action, which takes the following form
	\ie
	S_{\rm eff}=S_{\cD_{\rm IR}} +S_{\rm dilaton}[\tau]
	+\int_\Sigma d^4 z \sqrt{|h|}  \sum_{\cO_{\rm IR}\in \cD_{\rm IR}} \Omega^{4-\Delta_{\cO_{\rm IR}}}\lambda_{\cO_{\rm IR}}  \cO_{\rm IR}(z)
	\,.
	\label{SeffIR}
	\fe
	Here $S_{\cD_{\rm IR}}$ describes abstractly the IR DCFT and $S_{\rm dilaton}[\tau]$ is the dilaton effective action coming from integrating out massive modes along the flow. The coupling between $\cD_{\rm IR}$ and the dilaton $\tau$ is contained in the last term above, which is controlled by irrelevant defect operators $\cO_{\rm IR}(z)$ in the IR DCFT.
	
	Comparing UV and IR asymptotics of the RG flow, the anomaly matching condition boils down to
	\ie
	\D_\sigma \log  Z_{\cD_{\rm UV }}  =\D_\sigma \log Z_{\cD_{\rm IR}}+i\D_\sigma  S_{\rm dilaton}[\tau]\,,
	\fe
	which requires
	\ie
	\D_\sigma  S_{\rm dilaton}[\tau]=\int_\Sigma d^4x \sqrt{|h|} \sigma \,\Delta \cA^{\rm W}\,,
	\label{varStau}
	\fe
	with
	\ie
	\Delta \cA^{\rm W}\equiv \cA_{\cD_{\rm UV}}^{\rm W}-\cA_{\cD_{\rm IR}}^{\rm W}={1\over (4\pi)^2}(-\Delta a E_4+\Delta c W) + \Delta I_{\rm ext}(C^{(d)},K)\,,
	\label{ddta}
	\fe
	from \eqref{dta}, placing strong constraints on the dilaton effective action.
	
	The solutions for $S_{\rm dilaton}[\tau]$ can be found similar to the analysis in \cite{Komargodski:2011vj}. In general it takes the form
	\ie
	S_{\rm dilaton}[\tau] = S_{\rm WZ}[\tau] + S_{\rm inv}[\tau]\,.
	\label{Sdil}
	\fe
	The first term on the RHS is a \textit{cohomologically} nontrivial solution to the Wess-Zumino consistency condition and commonly referred to as the Wess-Zumino (WZ) term, which can be obtained from integrating the anomaly \cite{Wess:1971yu}
	\ie
	S_{\rm WZ}=&-\int_0^1 dt \int d^4 z \sqrt{|h|} e^{-4t\tau(z)}\tau \Delta\cA^{\rm W} (e^{-2t\tau(z)}h)\,.
	\fe
	We will be interested in the dilaton effective action for a flat defect on flat space, in which case we set $g_{\m\n}=\eta_{\m\n}$  and split the spacetime coordinates into tangential and orthogonal coordinates to the defect as $x^\m=(z^a,y^i)$.

	For the intrinsic defect conformal anomalies in \eqref{ddta}, the integral was performed in \cite{Schwimmer:2010za} which gives
	\ie
	S^{\rm int}_{\rm WZ}[\tau]=&\int_\Sigma  d^4 z \sqrt{|h|}\,\bigg( 
	-{\Delta a\over (4\pi)^2} \left(\tau E_4 + 4 \left (R^{ab}-{1\over 2} h^{ab} R\right)\pa_a \tau \pa_b \tau-4(\pa\tau)^2 \DA \tau +2(\pa\tau)^4  
	\right)
	\\
	&
	+{\Delta c\over (4\pi)^2}  \tau W
	\bigg)\,,
	\fe
	and reduces to the following simple form in flat space
	\ie
	S^{\rm int}_{\rm WZ}[\tau] \stackrel{\rm flat}{=} &  { \Delta a\over 8\pi^2}  \int_\Sigma  d^4 z\left(2(\pa \tau)^2 \DA \tau -(\pa\tau)^4  \right)\,.
	\label{SWZf}
	\fe
	The extrinsic anomalies on the other hand do not contribute to $S_{\rm WZ}$ on flat space. Under a Weyl transformation, the bulk Weyl curvature and the extrinsic curvature transform as
	\ie
	C^{(d)}_{\m\n\rho\sigma} \to C^{(d)}_{\m\n\rho\sigma} e^{-2\sigma}\,,\quad K^\m_{ab} \to K^\m_{ab}+ h_{ab}  N^{\m\n} \pa_\n\sigma\,.
	\label{wtck}
	\fe
	Here $N^\n{}_\m$ is the projector to the normal directions of $\Sigma$ defined by
	\ie
	g^{\m\n}=N^{\m \n}+ P^{\m\n}\,,\quad P^{\m\n}\equiv \pa_a X^\m \pa_b X^\n h^{ab}\,,
	\fe
	and $P^\m{}_\n$ is the projector to the tangential directions of $\Sigma$. In the flat space with $\sigma=t\tau(z)$ in \eqref{wtck}, we see clearly $\Delta I_{\rm ext}(C^{(d)},K)$ does not contribute to a WZ term. 
	
	The dilaton effective action may also contain terms that are Weyl invariant, corresponding to homogeneous solutions of \eqref{varStau}. They are captured by the second term $S_{\rm inv}[\tau]$ in \eqref{Sdil}. These terms can be constructed from the Weyl invariant worldvolume metric 
	\ie
	\hat h_{ab} \equiv e^{-2\tau} h_{ab}\,,
	\fe
	using the corresponding curvature invariants as in  \cite{Komargodski:2011vj,Komargodski:2011xv}. Such invariants again separate into intrinsic and extrinsic types. For similar reasons as explained above for the  WZ term, the extrinsic Weyl invariants do not play a role for the flat defect in flat space. With this understanding, we focus on the intrinsic invariants given by\footnote{Weyl invariant couplings between relevant operators in the IR DCFT and the dilaton of the form 
		\ie
		\int d^4 z \sqrt{|\hat h|}\,\Omega^{-\Delta_{\cO_{\rm IR}}}   \cO_{\rm IR},\quad   \int d^4 z \sqrt{|\hat h|}\,\Omega^{-\Delta_{\cO_{\rm IR}}}  \hat R \cO_{\rm IR} \,,
		\fe
		with $\Delta_{\rm IR}<4$ and $\Delta_{\rm IR}<2$ respectively must be tuned away otherwise it contradicts with the assumption of the flow into the IR DCFT \cite{Luty:2012ww}.
	}
	\ie
	S_{\rm inv}[\tau] =\int d^4 z \sqrt{|\hat h|}\,\left(
	\B_0+\B_1\hat R + \B_2\hat R^2 + \B_3\hat E_4 +\B_4 \hat C_{abcd} \hat C^{abcd} +
	\cO(\pa^6)
	\right)\,.
	\fe
	In the flat space limit, it gives the following interactions for $\tau$ up to the fourth derivative order,
	\ie
	S_{\rm inv}[\tau] \stackrel{\rm flat}{=}\int d^4 z  \left(\B_0 \Omega^4+
	6\B_1(\pa \Omega)^2+ 36\B_2\left(\DA \tau -(\pa \tau)^2 \right)^2
	+\cO(\pa^6)
	\right)\,.
	\label{Sinvf}
	\fe
	Putting everything together, from \eqref{SWZf} and \eqref{Sinvf}, we have the full defect dilaton effective action on flat space up to fourth derivative order which takes the identical form as in \cite{Komargodski:2011vj,Komargodski:2011xv}. After a redefinition \cite{Komargodski:2011xv},
	\ie
	\Psi\equiv 1-\Omega=1-e^{-\tau},
	\label{psidf}
	\fe
	we have 
	\ie
	S_{\rm dilaton}[\tau]
	\stackrel{\rm flat}{=}&
	\int d^4 z  \bigg(
	\B_0 (1-\Psi)^4+
	6\B_1(\pa \Psi)^2+ 36\B_2 {(\DA \Psi)^2\over (1-\Psi)^2}
	\\
	\quad & - { \Delta a\over 8\pi^2} \left({2 (\pa \Psi)^2 \DA \Psi \over (1-\Psi)^4}+ {(\pa \Psi)^4\over (1-\Psi)^4} \right)
	+\cO(\pa^6)
	\bigg)\,.
	\label{Sdff}
	\fe
	To prove the defect $a$-theorem, we would like explore unitarity constraints on the dilaton interaction proportional to $\Delta a$. This can be achieved by studying the four-point amplitude $A_4(s,t,u)$ of the probe dilaton $\Psi(p_1)\Psi(p_2)\to \Psi(-p_3)\Psi(-p_4)$ with external defect momenta $p_{i}$ and $s=-(p_1+p_2)^2,t=-(p_1+p_3)^2,u=-(p_1+p_4)^2$ are the usual Mandelstam variables. 
	To isolate the contribution from $\Delta a$, we work with the special kinematics such that the background dilaton is ``on-shell''
	\ie
	\DA \Psi=0\,.
	\fe
	In this case, the dilaton interactions in \eqref{Sdff}  simplify drastically. In particular the dilaton amplitude $A_4(s,t)$ at the fourth derivative order is completely determined by the interaction 
	\ie
	S_{\rm dilaton}[\tau] \supset  -{\Delta a\over 8\pi^2}\int d^4 z  (\pa \Psi)^4 \Rightarrow  A_4(s,t) \supset    {\Delta a\over 4\pi^2} (s^2+t^2+u^2)\,.
	\label{d4dt}
	\fe
	Furthermore we focus on the forward limit $p_1+p_3=p_2+p_4=0$, in which case the amplitude is a function of $s$ only and has the following small $s$ expansion,
	\ie
	A_4(s)={\rm const}+{\Delta a\over 2\pi^2}  s^2 +\cO(s^{\Delta_{\cO_{\rm IR}}-2})\,.
	\fe
	The corrections coming from the coupling between the IR defect $\cD_{\rm IR}$ and defect dilaton $\tau$ though the irrelevant operator $\cO_{\rm IR}$ in \eqref{SeffIR} are suppressed by $s^{\Delta_{\cO_{\rm IR}}-2}$.
	
	We expect the amplitude $A_4(s)$ to satisfy the Mandelstam analyticity of four-dimensional amplitudes, namely it is analytic on the upper half plane (from causality) and has branch cuts along the real axis due to the massless states in the IR DCFT that are being exchanged. The positivity of $\Delta a$ comes from a contour argument and standard dispersion relations for amplitudes.
	
	\begin{figure}[!htb]
		\centering
		\includegraphics[scale=1.3]{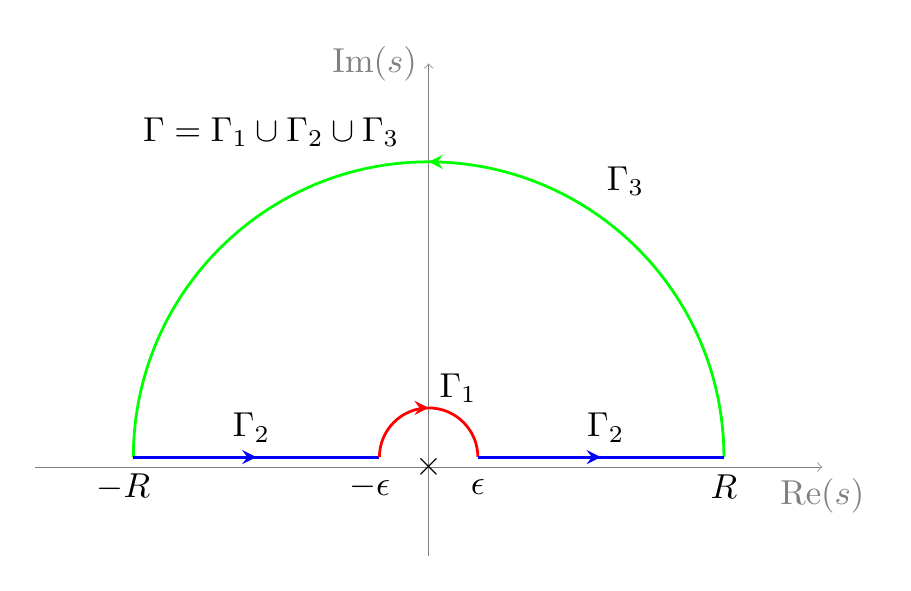}
		\caption{The integration contour $\Gamma=\Gamma_1\cup \Gamma_2\cup \Gamma_3$ for the dispersion argument. Here the limits $\ep \to 0$ and $R\to \infty$ are implicit.}
		\label{fig:dct}
	\end{figure}
	
	We start by considering the integral over a contour $\Gamma$ in the upper-half complex $s$-plane as in Fig~\ref{fig:dct}, 
	\ie
	0=-\pi i \oint_\Gamma ds {\pa_s A_4(s)\over s^2} =I_{\Gamma_1}+I_{\Gamma_2}+I_{\Gamma_3}
	\fe
	which vanishes by analyticity. The contour $\Gamma$ consists of three parts $\Gamma_{1},\Gamma_2,\Gamma_3$ and the integral splits accordingly as above. The part over the small semi-circle $\Gamma_1$ gives,
	\ie
	I_{\Gamma_1}=-\Delta a \,.
	\fe
	The integral over $\Gamma_2$ above the branch cuts yields,
	\ie
	I_{\Gamma_2}=2\pi \int_\ep^\infty ds {\pa_s {\rm Im} A_4(s)\over s^2}  \,,
	\fe
	where we have used crossing symmetry $A(s)=A(-s)$ since $u=-s$ in the forward limit, and the reality property of S-matrix $A_4(s^\star)=(A_4(s))^\star$ to combine the integrals over the two segments of $\Gamma_2$. Finally the integral over the large semi-circle $\Gamma_3$ at infinity vanishes
	\ie
	I_{\Gamma_3}=0\,,
	\fe
	since the large $s$ behavior of $A_4(s)$ is dominated by the least relevant UV perturbation in \eqref{UVdS} which gives $s^{\Delta_{\cO_{\rm UV}}-2}$ for $\Delta_{\cO_{\rm UV}}<4$.\footnote{A more careful argument shows that the same is true for marginally relevant deformations \cite{Luty:2012ww}.} 
	
	Therefore we arrive at the formula
	\ie
	\Delta a=2\pi\int_\ep^\infty ds {\pa_s {\rm Im} A_4(s)\over s^2}=4\pi\int_\ep^\infty ds { {\rm Im} A_4(s)\over s^3}\,,
	\label{ImA}
	\fe
	where the second equality comes from an integration-by-parts, which is possible thanks to the convergence properties of $A_4(s)$ (and thus its imaginary part) in the IR (${\rm Im} A_4(s)\sim s^{>2}$) and UV (${\rm Im} A_4(s)\sim s^{<2}$) asymptotics. These properties also ensure that the last integral in \eqref{ImA} is manifestly convergent. 
	
	Applying the optical theorem to the forward scattering $\Psi(p_1)\Psi(p_2)\to \Psi(p_1)\Psi(p_2) $ on the defect,\footnote{
		See \cite{Marino:2014oba,Heydeman:2020ijz} for recent discussions of the optical theorem for boundary field theories (and more general nonlocal field theories).} 
	\ie
	\Delta a={ 4\pi }\int_\ep^\infty ds {   s\sigma(s)\over s^3}\,,
	\fe
	where $\sigma(s)$ is the total cross-section for the scattering $\Psi(p_1)\Psi(p_2) \to \cD_{\rm IR}$ which involve DCFT states both on the defect and in the bulk. Unitarity of the defect field theory requires the above integrand to be positive. Since the integral is convergent, we arrive at the desired inequality
	\ie
	\Delta a >0\,,
	\fe
	that establishes Theorem~\ref{thm:athm}.
	
	\subsection{Explicit boundary RG flow in the free scalar theory}
	Here we discuss an explicit defect RG flow, between the Neumann and Dirichlet boundary conditions of the $d=5$ free scalar theory. In particular we will derive the nontrivial four-derivative interaction for the defect dilaton \eqref{d4dt} which captures the change $\Delta a$ in the defect conformal $a$-anomalies.
	
	The free scalar action on half space $\mR^{4,1}_+$ reads
	\ie
	S={1\over 2}\int_{y\geq 0} d^5 x \pa_\m \Phi \pa^\m \Phi \,,
	\fe
	with a boundary $\Sigma$ at $y=0$. 
	We start with the Neumann boundary condition 
	\ie
	\left. \pa_y \Phi \right|_\Sigma=0\,,
	\fe
	which admits a relevant deformation (boundary mass term) of the form
	\ie
	S_\Sigma = -{1\over 2} \int d^4 z \, m\Phi^2\,.
	\fe
	By varying the full action $S+S_\Sigma$ with respect to $\Phi$, we see the boundary condition gets deformed to
	\ie
	\left.\pa_y \Phi  -m \Phi \right |_\Sigma=0\,.
	\label{fsgb}
	\fe
	which flows to the Dirichlet boundary condition $\Phi|_\Sigma=0$ as $m \to \infty$.
	
	As explained in the last section, we introduce the defect dilaton $\tau$ to restore the defect conformal symmetry,
	\ie
	S_{\rm tot}[\tau]={1\over 2}\int_{y\geq 0} d^5 x \pa_\m \Phi \pa^\m \Phi -{1\over 2} \int_\Sigma d^4 z \, m \Phi^2+{1\over 2} \int_\Sigma d^4 z \, m \Psi \Phi^2\,,
	\label{fsfa}
	\fe
	with $\Psi$ defined as in \eqref{psidf}. By integrating out $\Phi$, we obtain the effective action for $\Psi$. Treating the last term in \eqref{fsfa} as a perturbation, the propagator for $\Phi$ subjected to the boundary condition \eqref{fsgb} can be found straightforwardly. We define the propagator after a Fourier transformation in the defect coordinates $z^a$ as
	\ie
	\hat G(p,-p,y,y') \equiv \la \Phi(p,y) \Phi(-p,y')\ra\,,
	\fe
	then the boundary condition \eqref{fsgb} requires
	\ie
	\left. (\pa_y-m) \hat G(p,-p,y,y')\right|_{y\to 0^+}=0\,.
	\fe
	We will work in the Euclidean signature and Wick rotate back to Minkowski signature later.
	The solution to the equation of motion and consistent with the boundary condition is
	\ie
	\hat G(p,-p,y,y')=
	{e^{-p|y-y'|}\over 2p} +{|p|-m\over  |p|+m } {e^{-p(y+y')}\over 2|p|}\,.
	\fe
	When the $\Phi$'s are restricted to boundary (which is what we need for computing the defect dilaton effective action), the propagator is simply
	\ie
	\hat G(p,-p,0,0)=
	{1\over |p|+m}\,.
	\fe
	
	To extract the four-point interaction of $\Psi$ in the IR effective action, we need to evaluate the following boundary one-loop Feynman diagrams in the large $m$ limit,
	\ie
	F_4(p_1,p_2,p_3,p_4)\equiv &\int   \prod_{i=1}^4 \left ( d^4 \vec z_i e^{i \vec p_i \cdot \vec  z_i} \right)\la  \Phi^2(  z_1,0) \Phi^2( z_2,0)\Phi^2( z_3,0)\Phi^2(z_4,0)\ra_c 
	\\
	=&\D^4(p_1+p_2+p_3+p_4) (I_{1234}+I_{1342}+I_{1423})\,,
	\fe
	with
	\ie
	I_{1234}\equiv & \int {d^4k \over (2\pi)^4}
	{1\over (|k|+m) (|k+p_1|+m)(|k+p_1+p_2|+m)( |k-p_4|+m)}\,,
	\fe
	and its cyclic permutations $I_{1342}$ and $I_{1423}$. In Appendix~\ref{app:integral}, we perform the integral explicitly and obtain
	\ie
	\left.F_4(p_1,p_2,p_3,p_4)\right|_{p_i^2=0}=\D^4(p_1+p_2+p_3+p_4) \left( {17\over 92160\pi^2}{s^2+t^2+u^2\over m^4} +\cO\left({s^4\over m^8},{t^4\over m^8}\right) \right)\,,
	\fe
	which corresponds to a dilaton interface of the form
	\ie
	S_{\rm dilaton} \supset -{2\over (4\pi)^2}{17\over 23040} \int d^4 z (\pa \Psi)^4 \,.
	\fe
	Compared to \eqref{Sdff}, we find that for the boundary RG flow in the free scalar theory
	\ie
	\Delta a={17\over 23040}\,.
	\label{caDN}
	\fe
	The boundary $a$-anomalies for the free scalar were also computed from heat kernel methods \cite{LEVITIN199835,Rodriguez-Gomez:2017aca} (see also \cite{Vassilevich:2003xt} for a review on these methods)\footnote{The defect conformal $a$-anomaly defined here is related to that in  \cite{Rodriguez-Gomez:2017aca} by $a_{\rm there}=-{2}a_{\rm here}$.}
	\ie
	a_{\rm Dir}=-{17\over 46080},\quad a_{\rm Neu}={17\over 46080}\,,
	\fe
	in agreement with what we found above. In Section~\ref{sec:5dbdyfree}, we will give a simple  rederivation of the above.

	\section{Defect Anomalies and $a$-Maximization}
	\label{sec:daamax}
	
	Conformal anomalies are important observables in CFTs but are generally hard to access in interacting theories of dimension $d>2$. This is partly because unlike in two dimensions where the conformal anomaly is simply determined by the two-point function of stress-tensor $T_{\m\n}$, the conformal anomalies in higher dimensions $d\geq 4$ are generally hidden in more complicated three- and higher-point functions of $T_{\m\n}$. Alternatively one can in principle determine the conformal anomaly of a CFT by computing its partition functions on curved backgrounds. However this is not feasible in practice for a general CFT. For conformal defects (or DCFTs), we face an even harder challenge, as we have little understanding for either of the two perspectives.
	
	In superconformal theories (SCFT), both obstacles can be overcome thanks to the preserved supersymmetry (SUSY). On the one hand, SUSY relates stress-tensor correlation functions to the simpler correlators that involve the conserved currents for R-symmetries, and consequently the conformal anomalies to the (mixed) R-symmetry 't Hooft anomalies. The latter are much easier to determine owing to their topological nature. On the other hand, the localization method allows one to determine the partition function exactly when sufficient supercharges are preserved (see \cite{Pestun:2016zxk} for a review). These methods  made possible the determination of conformal anomalies for a variety of SCFTs in $d=2,4,6$ \cite{Anselmi:1997am,Shapere:2008zf,Intriligator:2003jj,Benini:2012cz,Benini:2013cda,Cordova:2015fha,Cordova:2015vwa,Cordova:2019wns}
	and have recently been extended to superconformal surface defects \cite{Bianchi:2019sxz,Chalabi:2020iie,Wang:2020xkc}. Here we will carry out the analogous analysis for $p=4$-dimensional defects, providing a non-perturbative tool to access the defect conformal $a$- and $c$-anomalies for defects preserving the minimal supersymmetry.  
	
	These are defects invariant under an $\cN=1$ superconformal subalgebra of the full superconformal symmetry of the bulk SCFT,\footnote{The detailed subalgebra embeddings can be found in \cite{nw2}.} 
	\ie
	\mf{su}(2,2|1) \supset \mf{so}(4,2) \times \mf{u}(1)_R\,,
	\label{defectalg}
	\fe
	which contains four Poincar\'e supercharges and four conformal supercharges. The bosonic subalgebra generates the conformal subgroup longitudinal to the defect and the $U(1)_R$ symmetry that generally comes from a combination of the R-symmetry in the ambient SCFT and the transverse rotation symmetry (in $d=6$ only). In this section, we will derive the following universal relations between the defect conformal anomalies $a$ and $c$, and the defect 't Hooft anomalies $k_{RRR}$ and $k_R$ that involve the defect $U(1)_R$ symmetry (see \eqref{dJa}). 
	
	\begin{theorem}\label{thm:ack} 
		The conformal $a$- and $c$-anomalies of a four-dimensional $\cN=1$ superconformal defect is completely determined by the coefficients $k_{RRR}$ and $k_R$ for the $U(1)_R$ and mixed $U(1)_R$-gravity 't Hooft anomalies,  
		\ie
		a={9k_{RRR}-3k_R\over 32},\quad c={9k_{RRR}-5k_R\over 32}\,.
		\label{ackrel}
		\fe
	\end{theorem}
	These relations are identical to those satisfied by standalone 4d $\cN=1$ SCFTs \cite{Anselmi:1997am}, even though the nature and origin of the anomalies are very different. Furthermore, we will prove a defect version of the $a$-maximization principle \cite{Intriligator:2003jj} that identifies the superconformal $U(1)_R$ symmetry through a simple algebraic procedure. Together, they provide a powerful non-perturbative tool to extract the defect conformal anomalies in strongly coupled systems. 
	
	\subsection{Defect 't Hooft anomalies}
	We start by describing the defect 't Hooft anomalies associated to the defect $U(1)_R$ symmetry. In contrast with the case of a local 4d CFT, the Noether current $J^\m$ for the defect $U(1)_R$ symmetry is a $d$-dimensional conserved current, which satisfies the operator equation
	\ie
	\pa_\m J^\m(x)=0\,,
	\label{dJ0}
	\fe
	in flat space, everywhere including at the defect along $\Sigma$, but away from other operator insertions. This ensures that the $U(1)_R$ charge, defined by an integral of the current flux through a codimension-one submanifold $\cS$,
	\ie
	Q\equiv \int_\cS \star J\,,
	\fe
	remains topological when $\cS$ intersects with the defect $\Sigma$ (so that the charge $Q$ acts on the defect modes). 
	
	In the presence of other operator insertions, the current conservation law can be modified by contact-terms representing a (mixed) anomaly for the $U(1)_R$ symmetry. While this modification can occur both as contact terms in the bulk (for even $d$) and on the defect volume, here we will focus on the defect contributions. Upon coupling the bulk-defect system to background metric $g_{\m\n}$, $U(1)_R$ gauge field $A_\m$ and  gauge fields $B^i_a$ for additional abelian global symmetries $U(1)_i$ generated by locally conserved currents $J^i_a$ on the defect, the anomalous Ward identity for the $U(1)_R$ current can take the following general form\footnote{The abelian gauge fields $A$ and $B_i$ are anti-Hermitian in this paper, which differs by a factor of $i$ from those in \cite{Intriligator:2003jj}.}
	\ie
	\la \nabla_\m J^\m \ra_\cD \supset {\D(\Sigma)\over 2(2\pi)^2} \star_\Sigma \left( -{1\over 3}k_{ RRR} F\wedge F+ {1\over 24} k_{R} \tr \cR\wedge \cR 
	-  k_{Rij} F_i\wedge F_j
	-k_{RR i} F\wedge F_i
	\right) \,,
	\label{dJa}
	\fe
	with $F\equiv dA$ and $F_i \equiv dB_i$. The sums over the repeated $i,j$ indices are implicit. 
	Equivalently, the partition function of the defect field theory develops the following anomalous variations under a gauge transformation $A \to A+d\lambda$,
	\ie
	\D_\lambda \log Z_{\cD}= {1\over  2(2\pi)^2}
	\int_\Sigma d^4z\, \lambda   
	\left( -{1\over 3}k_{ RRR} F\wedge F+ {1\over 24} k_{R} \tr \cR\wedge \cR 
	-  k_{Rij} F_i\wedge F_j
	-k_{RR i} F\wedge F_i
	\right)
	\,,
	\label{vJa}
	\fe
	which is captured by the Stora-Zumino anomaly descent procedure \cite{Zumino:1983ew,AlvarezGaume:1984dr,Manes:1985df} from the following degree six anomaly polynomial \ie
	\cI_6=&-{k_R\over 24} c_1(F)p_1(T)+{k_{RRR}\over 6} c_1(F)^3+{k_{Rij}\over 2}c_1(F) c_1(F_i)c_1(F_j)
	+{k_{RRi}\over 2}c_1(F)^2c_1(F_i)
	\\
	&-{k_i\over 24} c_1(F_i)p_1(T)+{k_{ijk}\over 6} c_1(F_i)c_1(F_j)c_1(F_k)\,,
	\label{genAP}
	\fe
	where we have also included general anomaly terms involving the flavor symmetry.
	Here the characteristic classes are defined as usual,
	\ie
	c_1(F)={i\over 2\pi}  F,\quad p_1(T)=-{1\over 2 (2\pi)^2 }\tr(\cR\wedge \cR)\,.
	\fe
	The same anomaly polynomial \eqref{genAP} implies that the $p$-dimensional defect flavor symmetry currents $J^i_a$ receives an anomalous divergence
	\ie
	\la \nabla_a J^a_i(z) \ra_\cD \supset {1\over 2(2\pi)^2} \star_\Sigma \left( 
	-  k_{Rij} F\wedge F_j
	-k_{RR i} F\wedge F
	+ {1\over 24} k_{i} \tr \cR\wedge \cR 
	+ {1\over 3} k_{ijk} F_j\wedge F_k
	\right) \,,
	\label{dFa}
	\fe
	and the partition function varies accordingly under a gauge transformation $B^i \to A^i +d\omega^i$ on $\Sigma$
	\ie
	\D_{\omega^i} \log Z_{\cD}= {1\over  2(2\pi)^2}
	\int_\Sigma  d^4z \,\omega^i
	\left( 
	-  k_{Rij} F\wedge F_j
	-k_{RR i} F\wedge F
	+ {1\over 24} k_{i} \tr \cR\wedge \cR 
	+ {1\over 3} k_{ijk} F_j\wedge F_k
	\right)
	\,.
	\label{VFa}
	\fe
	
	\subsection{The supersymmetric anomaly multiplets}
	Supersymmetry generally leads to constraints on the anomalies admissible in a given supersymmetric (defect) field theory. In particular, for a superconformal defect $\cD$ in an ambient SCFT, since the stress-tensor $T_{\m\n}$ and the preserved $R$-symmetry current $J_\m$ are related by acting with the preserved supercharges $\cQ$, one naturally expects relations between the trace anomaly $T^\m_\m$ and the 't Hooft anomaly for the $R$-symmetry current, which define \textit{supersymmetric anomaly multiplets}. In the following we will establish the relations \eqref{ackrel} for $p=4$-dimensional $\cN=1$ superconformal defects.

	We will look for the supersymmetric completion of the anomalous Ward identity \eqref{dJa} and
	\ie
	\la T^\m_\m \ra_\cD \supset {1\over (4\pi)^2}\left (-a E_4 +c W \right)\,,
	\fe
	or equivalently that of the anomalous variation \eqref{vJa} and
	\ie
	\D_\sigma \log Z_\cD \supset {i\over (4\pi)^2}\int d^4 z \sqrt{|h|} \sigma \left (-a E_4 +c W \right)\,.
	\label{vTa}
	\fe
	We focus on the terms involving background metric and $U(1)_R$ gauge field.\footnote{The part of the anomaly that depends on the background flavor symmetry gauge fields has a separate supersymmetric completion.} The SUSY completion of \eqref{vJa} and \eqref{vTa} can be obtained by coupling to off-shell $\cN=1$ supergravity on the defect worldvolume $\Sigma$ and imposing the Wess-Zumino consistency conditions involving the supersymmetry, $U(1)_R$ and Weyl transformation generators (and their commutators) acting on $\log Z_\cD$. Because all these variations $\D_{(\cdot)} \log Z_\cD$ are local expressions on $\Sigma$, this problem is identical to the one solved in  \cite{Bonora:1984pn,Buchbinder:1986im} for standalone 4d theories, and gives rise to the super-Weyl anomaly for the DCFT. Here we follow the conventions of \cite{Buchbinder:1998qv} which was also used \cite{Schwimmer:2010za} except that our abelian gauge fields are anti-Hermitian.\footnote{The conventions of \cite{Buchbinder:1998qv} differ from those of \cite{Gates:1983nr} used in \cite{Anselmi:1997am} which employs a different set of torsion constraints.} The solution takes a simple form in the $\cN=1$ superspace with chiral and anti-chiral  Grassmann coordinates $\theta_\A,\bar \theta_{\dot \B}$. 
	The metric and R-symmetry gauge field are bosonic components of the supergravity superfield $\cH_{a}(z,\theta,\bar\theta)$, and the \textit{imaginary} R-symmetry gauge parameter $\lambda$ combines with the \textit{real} Weyl transformation parameter $\sigma$ into a chiral superfield $\D \Omega(z,\theta)$ satisfying
	\ie
	\D\Omega |_{\theta=0}=\sigma+{2\over 3}  \lambda\,.
	\fe
	The most general super-Weyl anomaly solving the Wess-Zumino consistency conditions is given by a chiral superspace integral together with its anti-chiral conjugate,
	\ie
	\D_\Omega \log Z_\cD[\cH_a]= 
	{1\over 2(4\pi)^2}
	\int_\Sigma d^4z d^2\theta \cE \D\Omega  \cA^{\rm SW}_{\cD}
	+(c.c)\,,
	\label{SWv}
	\fe
	where $\cE$ is the chiral superspace measure satisfying $ \cE |_{\theta=0}=\sqrt{|h|}$.
	The chiral anomaly density $\cA_\cD^{\rm SW}$ 
	is built from curvature superfields $W_{\A\B\C},G_a,R$ obtained from covariant super-derivatives $\cD_\A,\bar \cD_{\dot \A}$ acting on $\cH_a$, which contain the Weyl curvature, Ricci curvature and Ricci scalar respectively. Moreover $ G_a|_{\theta=\bar \theta=0}={4\over 3}i A_a$ is identified with the $U(1)_R$ gauge field. 
	The general solution (up to terms that are variations of diffeomorphism invariant local counter-terms),
	\ie
	\cA^{\rm SW}_{\cD}=\kappa_1 W^{\A\B\C}W_{\A\B\C} +\kappa_2\Xi 
	\label{ASW}
	\fe
	is a combination of the super-Weyl density $W^{\A\B\C}W_{\A\B\C}$ and the super-Euler density $\Xi$, which takes the following form in the old minimal supergravity\footnote{Unlike the super-Weyl density $W^{\A\B\C}W_{\A\B\C}$, the form of the super-Euler density $\Xi$ in terms of the superfields depends nontrivially on the chosen supergravity formulation (which is correlated with a choice of the supercurrent multiplet in the 4d field theory \cite{Komargodski:2010rb,Dumitrescu:2011iu}). See \cite{Bonora:2013rta,Assel:2014tba} for realizations of $\Xi$ in new minimal and non-minimal $\cN=1$ supergravities. }
	\ie
	\Xi\equiv   W^{\A\B\C}W_{\A\B\C}-{1\over 4} (\bar\cD^2-4R)(G^a G_a+2 R\bar R)\,.
	\fe
	To compare with the bosonic variations \eqref{vJa} and \eqref{vTa}, we need the F-term components of these composite chiral superfields (see \cite{Buchbinder:1998qv} and \cite{Schwimmer:2010za})\footnote{For example the first equality in \eqref{W2E4comp} comes from
		\ie
		\cD_\D W_{\A\B\C}=D_{(\A} W_{\B\C\D)}+{3\over 4} \ep_{\D (\A} \cD^\lambda W_{\B\C)\lambda}
		=D_{(\A} W_{\B\C\D)}+{3\over 4} \ep_{\D (\A} \cD^\lambda W_{\B\C)\lambda}\,,
		\fe
		and the following relations (only keeping terms dependent on metric and $U(1)_R$ gauge field)
		\ie
		D_{(\A} W_{\B\C\D)}|_{\theta=0}=(\sigma^{ab})_{\A\B}(\sigma^{cd})_{\C\D} C_{abcd}+\dots ,\quad D^\lambda W_{\A\B\lambda}|_{\theta=0}=i D_{(\A}{}^{\dot \A} G_{\B)\dot \A}|_{\theta=0}={4\over 3} (\sigma^{ab})_{\A\B} F_{ab}+\dots  \,,
		\fe
		from solving the torsion constraints and Bianchi identities (see Section 5.5.3 and 5.8.3 in \cite{Buchbinder:1998qv}). 
	}
	\ie
	\left.W_{\A\B\C}W^{\A\B\C} \right|_{\theta^2}
	\to\, &W-{8\over 3}F_{ab}F^{ab}+2i\star (\tr \cR\wedge \cR-{8\over 3} F\wedge F)\,,
	\\
	\left.\Xi \right|_{\theta^2}
	\to\, &E_4+2i\star (\tr \cR\wedge \cR- {40\over 9}F\wedge F)\,,
	\label{W2E4comp}
	\fe
	where we have dropped terms involving other supergravity fields on the RHS.
	
	Thus we have from \eqref{ASW} and \eqref{vTa}
	\ie
	c=-\kappa_1\,,\quad a=\kappa_2\,,
	\fe
	and from \eqref{ASW} and \eqref{vTa}
	\ie
	{k_{RRR}}={16\over 9} (3\kappa_1+5\kappa_2)\,,\quad k_R=-16(\kappa_1+\kappa_2)\,.
	\fe
	Together we arrive at \eqref{ackrel} as desired. 
	
	Next let us consider the SUSY completion of the 't Hooft anomaly \eqref{VFa} involving the 4d flavor symmetry current $J^i_a$. The gauge transformation parameter $\omega_i$ is promoted to a chiral superfield $\D\Lambda_i$ with $\omega_i=i {\rm Im} \,\D \Lambda_i |_{\theta=0}$\,. Focusing on the anomalous variations that depend only on $\D \Lambda_i$ and supergravity superfields, we have the SUSY completion\footnote{Note that an anomaly of the form $\int d^4 z d^2\theta \cE \D \Lambda_i \Xi+(c.c)$ is forbidden by the Wess-Zumino consistency condition $[\D_\sigma,\D_{\omega_i}]\log Z_\cD=0$ since we require the Weyl anomaly to be invariant under $U(1)_i$ gauge transformations.}
	\ie
	\D_{\Lambda^i} \log Z_\cD 
	={\kappa\over 4(2\pi)^2}\int_\Sigma d^4z d^2\theta  \cE  \D\Lambda_i W^{\A\B\C}W_{\A\B\C} +(c.c)\,.
	\fe
	Using \eqref{W2E4comp} and comparing with \eqref{VFa}, we arrive at the following relation between the mixed $U(1)_i$-$U(1)_R$ and $U(1)_i$-gravity anomalies, 
	\ie
	9 k_{RRi}=k_i=24\kappa\,.
	\label{flavorKrel}
	\fe

	\subsection{Defect $a$-maximization}
	Given a $p=4$-dimensional $\cN=1$ superconformal defect $\cD$, the 
	defect $U(1)_R$  symmetry is generally generated by a $d$-dimensional conserved current of the following form,
	\ie
	J_\m^t \equiv \hat  J_\m +  t_i \D(\Sigma) \D_\m^a J^i_a \,.
	\fe
	Here $\hat J_\m$ is a bulk current satisfying \eqref{dJ0} and (almost) determined by the embedding of the  defect superconformal symmetry $\mf{su}(2,2|1)$ in the bulk superconformal algebra. In particular, it is normalized such that the supercharges $\cQ$ preserved by the defect $\cD$ carries charges $\pm 1$.\footnote{The sign (and normalization) of $\hat J_\m$ is fixed by requiring its charge defined by $\hat R\equiv \int_{S^{d-1}} \star \hat J$ to appear in the anti-commutator of the supercharge $\cQ$ and its Hermitian conjugate in radial quantization as
		$\{\cQ,\cQ^\dagger\} \propto \Delta -{3\over 2}\hat R +\dots $\, where $\Delta$ is the usual dilatation operator.} The ambiguities come from locally conserved currents $J^i_a$ on the defect worldvolume $\Sigma$ with mixing coefficients $t_i$, whose symmetry charges commute with $\cQ$. 
	
	Following \cite{Intriligator:2003jj}, we define the trial conformal anomalies, in terms of the (mixed) 't Hooft anomalies involving the $U(1)_{R^t}$ symmetry generated by $J^t_\m$,
	\ie
	a(t)\equiv {9 k_{R^t R^t R^t}-3 k_{R^t}\over 32}\,,\quad c(t)\equiv {9 k_{R^t R^t R^t}-5 k_{R^t}\over 32}\,,
	\fe
	which coincide with \eqref{ackrel} for the genuine superconformal $U(1)_R$ symmetry at $t_i=t_i^\star$.
	Below we derive the defect version of the $a$-maximization principle that determines $t_i^\star$.
	
	\begin{theorem}[Defect $a$-Maximization]\label{thm:damax}
		The superconformal defect $U(1)_R$ symmetry 
		\ie
		J_\m= \hat  J_\m +  t^\star_i \D(\Sigma) \D_\m^a J^i_a
		\fe
		is determined by a local maximum $t_i=t^\star_i$ of the trial defect $a$-anomaly $a(t)$,
		\ie
		\left. \pa_i a(t) \right|_{t_i=t^\star_i}=0\,,\quad \left. \pa_i \pa_j a(t) \right|_{t_i=t^\star_i} <0\,.
		\label{amaxc}
		\fe
		Moreover the defect conformal anomalies are given by
		\ie
		a=a(t^\star)\,,\quad c=c(t^\star)\,.
		\fe
	\end{theorem}
	The first condition in \eqref{amaxc} simply follows from the anomaly multiplet relations for 't Hooft anomalies involving the $U(1)_i$ flavor symmetry \eqref{flavorKrel}. To derive the second condition in \eqref{amaxc}, we note that
	\ie
	\left. \pa_i \pa_j a(t) \right|_{t_i=t^\star_i} = {27\over 16}k_{Rij}\,.
	\fe
	In the following we will show that $k_{Rij}$ is negative definite as a consequence of unitarity and supersymmetry on the defect $\cD$.

	To explore possible constraints on the 't Hooft anomaly coefficient $k_{Rij}$, we turn on flavor symmetry background gauge fields $B^i_a$ and look for the SUSY completion of the corresponding anomaly terms in \eqref{vJa}
	\ie
	\D_\lambda \log Z_{\cD} \supset -{ k_{Rij}\over  2(2\pi)^2}
	\int_\Sigma d^4z\, \lambda   
	F_i\wedge F_j\,.
	\label{varRij}
	\fe
	Promoting $B^i_a$ to a background vector superfield on $\Sigma$ with field strength chiral superfield $W^i_\A$, we have
	\ie
	\D_\Omega \log Z_\cD \supset &
	{\kappa_{ij}\over (4\pi)^2}
	\int_\Sigma d^4z d^2\theta \cE \D\Omega   W_\A^i W^{\A j}
	+(c.c)
	\\
	=&
	{1\over (4\pi)^2}\int_\Sigma d^4z \sqrt{|h|} \sigma \left( {\rm Re}\kappa_{ij} F^i_{ab}F^{j ab} - {\rm Im}\kappa_{ij} \ep^{abcd} F^i_{ab}F^{j}_{cd}\right )
	\\
	&
	+
	\lambda
	\left( {\rm Im}\kappa_{ij} F^i_{ab}F^{j ab} + {\rm Re}\kappa_{ij} \ep^{abcd} F^i_{ab}F^{j}_{cd}\right )+\dots
	\,.
	\fe
	Comparing with \eqref{varRij}, we find\footnote{Here we have assumed the absence of an exotic parity-even anomaly of the type
		\ie
		\nabla_\m J^\m \sim \D(\Sigma) C_{ij} F^{i}_{ab}F^{jab} \,,
		\fe
		for the DCFT. This would be a defect analog of the \textit{impossible anomaly} discussed in \cite{Nakayama:2018dig}. 
	}
	\ie
	{\rm Im} \kappa_{ij}=0,\quad {\rm Re} \kappa_{ij}=-2k_{Rij}\,,
	\fe
	which implies a flavor contribution to the trace anomaly,
	\ie
	\la T^\m_\m \ra_\cD \supset  -2 \D(\Sigma)  k_{Rij} F^i_{ab} F^{j ab}\,.
	\label{TJJ}
	\fe
	Now recall the two-point functions of the conserved currents $J^i_a$ are fixed by conformal symmetry to take the form
	\ie
	\la J_a^i(z_1) J_b^j (z_2)\ra _\cD 
	=
	\tau^{ij}(\pa^2\D_{ab}-\pa_a\pa_b){1\over |z_{12}|^4}\,,
	\fe
	with positive definite coefficient $\tau^{ij}$ from unitarity. The RHS suffers from a short distance singularity which  can be regularized using\cite{Osborn:1993cr}
	\ie
	\cR \left ({1\over z^4} \right)= -{1\over 4} \DA {\log (\m^2 z^2)\over z^2}\,,
	\fe
	and the conformal anomaly arises from the dependence of the regularization scheme on the scale parameter $\mu$ as in \ie
	\m {\pa \over \pa \m} \cR \left ({1\over z^4} \right)=2\pi^2\D^4(z)\,.
	\fe
	Therefore we have
	\ie
	\m {\pa \over \pa \m}\la J_a^i(z_1) J_b^j (z_2)\ra_\cD =2\pi^2 \tau^{ij} (\pa^2\D_{ab}-\pa_a \pa_b) \D^4(z_1-z_2)\,.
	\fe 
	On the other hand,
	\ie
	\m {\pa \over \pa \m}\la J_a^i(z_1) J_b^j (z_2)\ra_\cD =\la \int_\cM d^5x\, T^\m_\m (x) J^i_a(z_1) J^j_b(z_2) \ra_\cD 
	=-2k_{Rij}(\pa^2\D_{ab}-\pa_a \pa_b) \D^4(z_1-z_2)\,,
	\fe
	where the last equality follows from \eqref{TJJ}. Thus we conclude that
	\ie
	k_{Rij}=-\pi^2 \tau_{ij} \,,
	\fe
	which is negative definite as desired. 
	
	Before ending this section, let us make a few comments on the $a$-maximization procedure and its relation to the defect $a$-theorem. The $a$-maximization holds with respect to all $U(1)$ flavor symmetry currents on the defect worldvolume $\Sigma$. In practice, we often do not directly deal with the strongly coupled fixed point. Instead we infer the set of the $U(1)$ symmetries from a nearby (Lagrangian) description, which we use to determine the relevant 't Hooft anomalies and obtain the trial $a$-function before maximizing. It can happen that there are accidental symmetries that are missed in this way, which may lead to nonsensical answers for the $U(1)_R$ symmetry and conformal anomalies (e.g. a naive violation of the unitarity bound for certain defect operators). In such cases, we have to adjust the ansatz for the candidate $U(1)_R$ symmetry by including the accidental symmetries (e.g. from operators that hit the unitarity bound) and redo the $a$-maximization (see \cite{Kutasov:2003iy} for relevant discussions in 4d SCFTs).
	
	As explained in \cite{Intriligator:2003jj}, the $a$-maximization principle almost implies the $a$-theorem for supersymmetric RG flows triggered by (marginally) relevant perturbations, since the maximization procedure is performed over a larger space of $U(1)$ symmetries in the UV than in the IR. This was later made more precise in \cite{Kutasov:2003iy,Barnes:2004jj} by constructing explicitly an $a$-function along the supersymmetric RG flow with the desired properties as in the strongest version of the $a$-theorem. A direct generalization of their construction  leads to the defect $a$-theorem for supersymmetric RG flows from (marginally) relevant defect perturbations. Once again, cases with accidental symmetries must be treated with care \cite{Barnes:2004jj}.

	\section{Defect $a$-anomalies in SCFTs}
	\label{sec:examples}
	
	$\cN=1$ superconformal defects of dimension $p=4$ exist in 5d and 6d SCFTs.\footnote{See \cite{Agmon:2020pde,nw2} for a general classification of unitary superconformal defects based on the preserved (and broken) symmetries.} 
	These SCFTs are generally strongly coupled and do not have conventional perturbative Lagrangian descriptions, which makes it especially challenging to study the defects thereof. 
	In the following, we will apply our non-perturbative methods developed in the last section to a number of examples and determine their defect conformal anomalies exactly. 
	
	\subsection{Boundaries for 5d $\cN=1$ SCFTs}
	In 5d $\cN=1$ SCFTs,  the $p=4$-dimensional superconformal defects appear either as half-BPS boundaries or interfaces. The defect $U(1)_R$ symmetry is identified with the Cartan generator of the bulk $SU(2)_R$ symmetry,
	\ie
	R=R_{5d}\,,
	\label{5dR}
	\fe
	up to mixing with flavor symmetry currents on the defect.
	A half-BPS interface between two 5d SCFTs $\cT_1$ and $\cT_2$ is related by the folding trick to a half-BPS boundary for the doubled theory $\cT_1\otimes \bar \cT_2$ (where the second factor undergoes an orientation flip). For this reason, we will focus on superconformal boundaries in 5d SCFTs. 
	
	\subsubsection{Boundary 't Hooft anomalies from bulk fermions}
	To determine the boundary conformal anomalies using our method requires the knowledge of the boundary 't Hooft anomalies involving the superconformal $U(1)_R$ symmetry. If the relevant DCFT admits a $U(1)_R$ preserving deformation to a free theory, one can hope to determine the 't Hooft anomalies from those of the free fields. In $d=5$, such boundary anomalies can come from bulk Dirac fermions (and complex two-forms).\footnote{See earlier works \cite{Horava:1995qa,Horava:1996ma} for discussions of anomaly inflow to the boundary from bulk massless fermions, in the context of the $E_8$ end-of-the-world brane in 11d supergravity.}

	Let us consider a 5d Dirac fermion $\Psi_{\rm Dirac}$ on half space $\mR^{4,1}_+$ with a timelike boundary $\Sigma$ at $y=0$, and suppose it has charge $q$ under a $U(1)$ global symmetry $\Psi_{\rm Dirac} \to e^{i q\theta} \Psi_{\rm Dirac}$ and also transform in an irreducible representation $\rho$ for an nonabelian global symmetry $G$. The standard $U(1)\times G$-preserving boundary conditions for $\Psi_{\rm Dirac}$ 
	on $\Sigma$ at $y=0$ 
	correspond to 
	\ie
	P_\pm\Psi_{\rm Dirac}|_\Sigma=0 \,,
	\label{fbc}
	\fe
	where $P_\pm \equiv {1\over 2}(1_4\pm \CC_y)$ is a (anti)chiral projector on the boundary.\footnote{Here $\CC_y=i\CC^0\CC^1\CC^2\CC^3$ coincides with the standard 4d chirality matrix (see Appendix~\ref{app:5dsusy} for the spinor conventions).} The boundary (mixed) 't Hooft anomalies involving the $U(1)$ symmetry are as summarized in Table~\ref{tab:bta}.
	
	\begin{table}[!htb]
		\centering
		\renewcommand{\arraystretch}{1.8}
		\begin{tabular}{|c|c|c|}
			\hline
			& Fields         & Anomaly $\cI_6$    \\ \hline
			\multirow{2}{*}{$4d$} &   $\psi$  & ${d_\rho\over 6}q^3 c_1(F)^3-{d_\rho\over 24}q p_1(T) c_1(F)
			- {T_\rho } q c_1(F)c_2(F_G)
			+ {a_\rho\over 6}c_3(F_G)
			$ 
			\\\cline{2-3}
			&  $\bar\psi$  & $-{d_\rho\over 6}q^3 c_1(F)^3+{d_\rho\over 24}q p_1(T) c_1(F)
			+ {T_\rho }q c_1(F)c_2(F_G)
			- {a_\rho\over 6}c_3(F_G)
			$  \\\hline
			\multirow{2}{*}{$5d$} &    $\left.P_+\Psi_{\rm Dirac}\right|_\Sigma=0$   & $-{d_\rho\over 12}q^3 c_1(F)^3+{d_\rho\over 48} q p_1(T) c_1(F)
			+ {T_\rho\over 2}q c_1(F)c_2(F_G)
			- {a_\rho\over 12}c_3(F_G)
			$ 
			\\\cline{2-3}
			&   $\left.P_-\Psi_{\rm Dirac}\right|_\Sigma =0$   & ${d_\rho\over 12}q^3 c_1(F)^3-{d_\rho\over 48} q p_1(T) c_1(F)
			- {T_\rho\over 2 }q c_1(F)c_2(F_G)
			+ {a_\rho\over 12}c_3(F_G)
			$  
			\\\hline    
		\end{tabular}
		\caption{The 't Hooft anomalies contributed by a 4d Weyl fermion $\psi$ and its conjugate $\bar\psi$, and a 5d Dirac fermion $\Psi_{\rm Dirac}$ with different boundary conditions. Both $\psi$ and $\Psi_{\rm Dirac}$ carry charge $q$ under the $U(1)$ symmetry and transform in an irreducible representation $\rho$ for the nonabelian $G$ symmetry. The dimension, Dynkin index and cubic Casimir eigenvalue for $\rho$ are denoted by $d_\rho$, $T_\rho$ and $a_\rho$ respectively. 
			The background gauge connections for $U(1)\times G$ are $A$ and $A_G$ respectively.}
		\label{tab:bta}
	\end{table}

	One way to see this is by turning on a $U(1)\times G$ preserving mass deformation in the bulk\footnote{See \cite{Dimofte:2017tpi} for similar discussions of a 3d Dirac fermion on half space $\mR^3_+$.}
	\ie
	S_{\rm Dirac}=\int_{\mR^{4,1}_+} d^5x\, \bar \Psi_{\rm Dirac} (\CC^\m D_\m - m) \Psi_{\rm Dirac}+{1\over 2}\int_\Sigma \bar \Psi_{\rm Dirac} \CC^y \Psi_{\rm Dirac}\,,
	\fe
	where $\bar \Psi_{\rm Dirac}\equiv \Psi^\dagger_{\rm Dirac} i \CC^0$ as usual and the boundary term is necessary for the reality of the action. 
	Integrating out the massive Dirac fermion in the bulk generates a 5d Chern-Simons term for the background $U(1)\times G$ gauge fields $A$ and $A_G$, and Riemann curvature 2-form $\cR$ in the effective action \cite{Witten:1996qb,Bonetti:2013ela},\footnote{One may be cautious about the unquantized Chern-Simons level here as it would not be gauge invariant under large gauge transformations. Here we emphasize that the relevant physical information is just contained in the infinitesimal gauge variation of  \eqref{bulkinflow} which is well defined, and can be verified, for example, by a direct calculation of the divergence of the 5d $U(1)$ current using Feynman diagrams for the fermions satisfying the boundary conditions \eqref{fbc}. Towards the end of this section, we will also provide other arguments that lead to the same conclusions for these boundary anomalies. Nonetheless it is certainly desirable to understand these anomalies for general bulk-defect coupled systems, from the modern perspective (see e.g. \cite{Freed:2014iua,Monnier:2019ytc}) using invertible field theories in one higher dimension, with suitable generalizations.
	}
	\ie
	-{{\rm sign}(m)\over 2}\int_{\mR^{4,1}_+} \left( {d_\rho q^3 \over 24 \pi^2 }A\wedge F\wedge F + {d_\rho q\over 192 \pi^2 } A\wedge \tr(\cR\wedge \cR) 
	- {T_\rho q \over 8 \pi^2 } A\wedge \Tr (F_G\wedge F_G)+ {a_\rho \over 6} {\rm CS}_5(A_G)
	\right)\,,
	\label{bulkinflow}
	\fe
	where the last term is the usual non-abelian Chern-Simons 5-form defined by $d{\rm CS}_5(A_G)=-2\pi c_3(F_G)$ and 
	the Chern-Simons level depends on the sign of $m$. In the above, $d_\rho$ is dimension of the representation $\rho$, $T_\rho$ is the Dynkin index and $a_\rho$ denotes the cubic Casimir eigenvalue. For $G=SU(N)$ and $\rho=\Box$ (the fundamental representation), $d_\rho=N$, $T_\rho={1\over 2}$ and $a_\rho=1$.
	
	The Chern-Simons action \eqref{bulkinflow} clearly contributes to the boundary 't Hooft anomalies through the inflow \cite{Callan:1984sa}, but we also need to remember there maybe residual massless boundary modes from the massive 5d fermion. Indeed, the equation of motion
	\ie
	(\CC^y\pa_y +m + \CC^a \pa_a ) \Psi_{\rm Dirac}=0\,,
	\fe
	implies that a normalizable boundary massless mode $\Psi_{\rm Dirac}(y)\sim e^{-|m|y}$ is possible if
	\ie
	\CC_y \Psi_{\rm Dirac}|_\Sigma  = {\rm sign}(m)  \Psi_{\rm Dirac} |_\Sigma \,,
	\label{zmreq}
	\fe
	whose contribution to the boundary 't Hooft anomaly comes from the inflow of 
	\ie
	{\rm sign}(m)\int_{\mR^{4,1}_+} 
	\left( {d_\rho q^3 \over 24 \pi^2 }A\wedge F\wedge F + {d_\rho q\over 192 \pi^2 } A\wedge \tr(\cR\wedge \cR) 
	- {T_\rho  q\over 8 \pi^2 } A\wedge \Tr (F_G\wedge F_G)+ {a_\rho \over 6} {\rm CS}_5(A_G)
	\right)\,,
	\label{bdyzm}
	\fe
	as ${\rm sign}(m)$ coincides with the chirality of the boundary massless fermion according to \eqref{zmreq}. It is now straightforward to verify the entries in Table~\ref{tab:bta} based on the above.
	For example, with the boundary condition $P_+\Psi_{\rm Dirac}|_\Sigma=0$, the full boundary anomaly is just given by \eqref{bulkinflow} with the overall coefficient $-{1\over 2}$ since for $m>0$ this is the only contribution. For $m<0$, a boundary massless chiral fermion is admissible by \eqref{zmreq} and the total contribution from  \eqref{bulkinflow} and \eqref{bdyzm} is identical as before. 
	
	One can also understand the relation between the boundary anomaly from a bulk fermion satisfying $P_-\Psi_{\rm Dirac}|_\Sigma=0$ and that of a 4d boundary chiral fermion $\psi$ as follows. Let us put the 5d Dirac fermion on a slab $\Sigma \times [0,L]$ with identical boundary conditions $P_-\Psi_{\rm Dirac}=0$ at the two ends $y=0$ and $y=L$. For small $L$ and in the low energy limit, this is the same as a chiral fermion on $\Sigma$. Thus the anomaly for a single boundary is half of that of a chiral fermion. Alternatively, starting with a boundary satisfying $P_+\Psi_{\rm Dirac}|_\Sigma=0$, we can couple the bulk fermion to a boundary chiral fermion $\psi$ by $\int_\Sigma d^4z\,\bar \Psi_{\rm Dirac} P_+\psi+(c.c)$. Integrating out $\psi$, this flips the boundary condition of the bulk Dirac fermion to $P_-\Psi_{\rm Dirac}|_\Sigma=0$. Since the boundary conditions $P_\pm\Psi_{\rm Dirac}|_\Sigma=0$ have opposite anomalies by parity, we again reach the same conclusion.

	\subsubsection{Boundary conditions for free hypermultiplets}
	\label{sec:5dbdyfree}
	We start by considering the simplest 5d SCFT defined by a free hypermultiplet which consists of four real scalars $\Phi^{i A}$ and a symplectic-Majorana fermion $\Psi^A$.\footnote{\label{foot:SMvD}The symplectic-Majorana condition is 
		\ie(\Psi_\A^A)^*=C^{\A\B} \ep_{AB} \Psi_\B^B\fe where $\A,\B=1,2,3,4$ are the spinor indices and $C^{\A\B}$ is the anti-symmetric 5d charge conjugation matrix. Consequently it captures the same independent degrees of freedom as a Dirac fermion $\Psi_{\rm Dirac}\equiv \Psi^1$.
	} The theory has an $SU(2)_R\times SU(2)_F$ symmetry and $i=1,2$ and $A=1,2$ are the corresponding doublet indices. We place the theory on half space $\mR^{4,1}_+$ and consider superconformal boundary conditions on $\Sigma$ at $y=0$.
	
	With regard to the boundary $\cN=1$ superconformal symmetry, the hypermultiplet splits into two chiral multiplets $(X,\psi_X)$ and $(Y,\psi_Y)$ on $\Sigma$,
	\ie
	X=q^{11}\,,\quad\psi_X= P_+\Psi^1\,,\quad 
	Y=q^{12}\,,\quad\psi_Y= P_+ \Psi^2\,.
	\fe
	The boundary superconformal $U(1)_R$ symmetry is identified the Cartan of $SU(2)_R$ symmetry as in \eqref{5dR}, under which the complex scalars $(X,Y)$ carry charge $+1$ but the fermions $(\psi_X,\psi_Y)$ are uncharged.
	
	The simplest supersymmetric boundary conditions come from putting together the Neumann and Dirichlet boundary conditions for the scalars, and the standard $U(1)_R$ symmetric boundary conditions for the fermions \eqref{fbc} \cite{Dimofte:2012pd},\footnote{Note that in terms of the Dirac fermion $\Psi^1$ we have $\psi_Y=(P_- \Psi^1)^*$. See Appendix~\ref{app:5dsusy} for further details.}
	\ie
	\cB_X[\Phi]:~&Y|_\Sigma  =\psi_Y|_\Sigma  =\pa_y \bar X|_\Sigma  =0\,,
	\\
	\cB_Y[\Phi]:~&X|_\Sigma  =\psi_X|_\Sigma  =\pa_y \bar Y|_\Sigma  =0\,,
	\label{BXBY}
	\fe
	which amounts to setting one of the two chiral multiplets to zero identically.
	The corresponding boundary conformal anomalies are determined by the 't Hooft anomalies as in \eqref{ackrel}. The latter can receive inflow contributions from charged fermions in the bulk as explained in the last section. Here the boundary 't Hooft anomalies vanish since the fermions are uncharged under $U(1)_R$. Consequently, the defect conformal anomalies all vanish, as in Table~\ref{tab:ndclcr}.

	In fact, the boundary conditions $\cB_{X,Y}[\Phi]$ are special points on a $\mathbb{CP}^1$ conformal manifold of superconformal boundaries for a free 5d hypermultiplet, with the same vanishing defect conformal anomalies. This boundary conformal manifold comes about from the bulk $SU(2)_f$ flavor symmetry acting on the boundary conditions \eqref{BXBY} which preserves a $U(1)_f$ subgroup.\footnote{In general, flavor symmetry currents broken by a conformal defect give rise to exactly marginal couplings for the defect \cite{Dimofte:2012pd,Agmon:2020pde}.} Close to the $\cB_{X}[\Phi]$ point, the $SU(2)_f$ rotation induces a marginal perturbation 
	\ie
	\D \cS_{\cB_X[\Phi]}=\zeta \int_\Sigma d^4 z d^2\theta X^2 + (c.c)\,.
	\fe
	More general superconformal boundaries for the 5d hypermultiplet can be obtained by coupling $\cB_X[\Phi]$ to a 4d $\cN=1$ SCFT $\cT_{4d}$ on $\Sigma$ through a superpotential, 
	\ie
	\cB_X[\Phi] \oplus \cT_{4d}~{\rm with}~\int_\Sigma d^4 z d^2\theta X \cO_{4d}+(c.c)
	\to 
	\cB_{\rm gen}[\Phi]\,,
	\label{XT4dcoup}
	\fe
	where $\cO_{4d}$ is a scalar chiral primary operator in $\cT_{4d}$ of $U(1)_R$ charge $R(\cO_{4d})\leq 1$ and scaling dimension $\Delta(\cO_{4d})={3\over 2}R(\cO_{4d})\leq {3\over 2}$. 
	
	If $R(\cO_{4d})=1$, the coupling in \eqref{XT4dcoup} is exactly marginal\footnote{This is because scalar chiral primary operators in 4d $\cN=1$ SCFTs are absolutely protected if $R<2$ and thus cannot develop anomalous dimensions \cite{Cordova:2016emh}.} and the total boundary conformal anomalies simply coincide with the anomalies of the 4d SCFT $\cT_{4d}$,
	\ie
	a(\cB_{\rm gen}[\Phi])=a(\cT_{4d})\,,\quad c(\cB_{\rm gen}[\Phi])=c(\cT_{4d})\,.
	\fe
	We emphasize that these are generally strongly coupled boundary conditions for the free hypermultiplet. In particular this includes the example of an $E_7$-invariant boundary condition for 28 free hypermultiplets, obtained by an exactly marginal coupling to the 4d $\cN=1$ $SU(2)$ SQCD with $N_f=4$ on the boundary \cite{Dimofte:2012pd}.

	If instead $R(\cO_{4d})<1$, the coupling between the boundary condition $\cB_X[\Phi]$ and the
	SCFT $\cT_{4d}$ is relevant and should flow to the new superconformal boundary $\cB_{\rm gen}[\Phi]$. In the simplest scenario, $\cT_{4d}$ is a free chiral multiplet $\phi$ of R-charge $2\over 3$ and we can take $\cO_{4d}=\phi$. Then the superpotential deformation $\int_\Sigma d^4 zd^2\theta \phi X$  simply imposes Dirichlet boundary condition on $X$ while lifting the Dirichlet boundary condition on $Y$ (by $Y=\phi$) and thus we have a boundary RG flow \cite{Dimofte:2012pd}
	\ie
	\cB_X[\Phi] \oplus \phi~{\rm with}~\int_\Sigma d^4 z d^2\theta X \phi+(c.c)
	\to 
	\cB_{Y}[\Phi]\,,
	\fe
	which is trivially consistent with the boundary $a$-theorem. We leave the investigation of more general boundary conditions for the free hypermultiplet that arise this way to future work.\footnote{Note that for the relevant coupling in \eqref{XT4dcoup} to preserve a manifest $U(1)_R$ symmetry, the SCFT $\cT_{4d}$ should have a $U(1)$ flavor symmetry under which the operator $\cO_{4d}$ is charged.}
	
	From the above discussion, it should be clear that the zoo of interacting superconformal boundary conditions for the free hypermultiplets is rather rich, and it would be interesting to classify them from the bootstrap approach along the lines of \cite{Behan:2020nsf}.
	
	Finally let us comment on our results in relation to the free field defect conformal anomalies obtained in \cite{Rodriguez-Gomez:2017aca} from heat kernel computations \cite{Vassilevich:2003xt}. The results for boundary $a$-anomalies are tabulated in Table~\ref{tab:ndclcr} and they are consistent with our findings.
	The fact that the Dirichlet and Neumann boundary conditions for a scalar $\Phi$ contribute opposite defect conformal anomalies is easy to understand. We take two scalars $\Phi_1$ and $\Phi_2$ on $\mR^{4,1}_+$ satisfying conformal Dirichlet and Neumann boundary conditions respectively, and then turn on an exactly marginal coupling given by $\int_\Sigma  d^4 z \Phi_2 \pa_y \Phi_1$. After unfolding, this coupling   identifies the two scalars $\Phi_1$ and $\Phi_2$ living on $\mR^{4,1}_+$ and $\mR^{4,1}_-$ respectively at $\Sigma$, and the original boundary corresponds to a transparent interface. Since the defect $a$-anomaly does not depend on marginal couplings on the defect due to the Wess-Zumino consistency conditions \cite{Wess:1971yu,Osborn:1991gm}, the original boundary must have vanishing total $a$-anomaly. Together with the difference $a_{\rm Neu}-a_{\rm Dir}$ computed in \eqref{caDN}, this gives a re-derivation of the results for $a_{\rm Neu}$ and $a_{\rm Dir}$ in \cite{Rodriguez-Gomez:2017aca}.
	
	Following a similar argument for two fermions $\Psi_1$ and $\Psi_2$ with boundary conditions $P_+\Psi_1|_\Sigma=P_-\Psi_2|_\Sigma=0$ and boundary marginal coupling $\int_\Sigma d^4 z \bar\Psi_1 \CC_y P_+\Psi_2$, we conclude that the total $a$-anomaly again vanishes. Here the chiral and anti-chiral boundary conditions are further related by a parity-reversal along $\Sigma$ which does not affect the boundary $a$- or $c$-anomalies which are parity-even. Therefore the individual boundary $a$-anomalies for $\Psi_1$ and $\Psi_2$ must vanish, again consistent with the explicit computations in \cite{Rodriguez-Gomez:2017aca}. 
	
	The situation is less clear for the boundary conformal $c$-anomalies. The free field boundary $c$-anomalies have not been computed to the author's knowledge. Nonetheless the fermion cases are restricted by the parity symmetry as in Table~\ref{tab:ndclcr},
	and the vanishing $c$-anomaly for the supersymmetric boundary $\cB_X[\Phi]$ requires
	\ie
	c_{\rm Dir}+c_{\rm Neu}+2c_\Psi =0\,.
	\fe
	The precise values of the individual $c$-anomalies above should be accessible from the bulk stress-tensor two-point function in the presence of the boundary (see related discussions in lower dimensions in \cite{Herzog:2017kkj}).

	\begin{table}[!htb]
		\centering
		\renewcommand{\arraystretch}{1.3}
		\begin{tabular}{|c|c|c|c|}
			\hline
			&   Fields        &  $a$ &  $c$  \\ \hline
			\multirow{5}{*}{$4d$}  &  real scalar  &  ${1\over 360}$ &  ${1\over 120}$
			\\\cline{2-4} &   Weyl fermion  & ${11\over 720}$ & ${1\over 40}$
			\\\cline{2-4} &   photon  & ${31\over 180}$ & ${1\over 10}$
			\\\cline{2-4}&   chiral  &  ${1\over 48}$ & ${1\over 24}$
			\\\cline{2-4}
			&   vector &   ${3\over 16}$ & ${1\over 8}$
			\\\hline
			\multirow{6}{*}{$5d$} &        $\left.\pa_y \Phi \right|_\Sigma =0$   & $ {17\over 46080}$  &   $c_{\rm Neu}$
			\\\cline{2-4}
			&   $\left. \Phi \right|_\Sigma =0$   &  $-{17\over 46080}$  & $c_{\rm Dir}$
			\\\cline{2-4}
			& $\left.P_+\Psi\right|_\Sigma=0$   & $0$  &  $c_\Psi$
			\\\cline{2-4}
			&   $\left.P_-\Psi\right|_\Sigma =0$   & $0$  &  $c_\Psi$
			\\\cline{2-4}
			
			&   $\cB_{X,Y}[\Phi]$   & $0$  &    0
			\\\hline    
		\end{tabular}
		\caption{The conformal anomalies of 4d free fields and $\cN=1$ supermultiplets, and the boundary conformal anomalies of 5d free fields and their supersymmetric completions.}
		\label{tab:ndclcr}
	\end{table}

	\subsubsection{Boundaries of $E_n$ SCFTs}
	Let us now discuss superconformal boundaries of interacting 5d $\cN=1$ SCFTs. A particularly well-studied set of examples known as the $E_n$ SCFTs for $0\leq n\leq 8$ originated from \cite{Seiberg:1996bd,Morrison:1996xf,Douglas:1996xp} and was generalized in \cite{Intriligator:1997pq}. For $n\geq 1$, the $E_n$ SCFT, upon a supersymmetric mass deformation, is described by an $\cN=1$  $SU(2)$ super-Yang-Mills theory coupled to $n-1$ hypermultiplets transforming in the fundamental representation. The 5d gauge theories are non-renormalizable and the $E_n$ SCFTs are expected to be their UV completions. The manifest  global symmetry in the IR gauge theory is $U(1)_I\times SO(2n-2)$ where the first factor comes from the topological instanton current and the second factor is due to the fundamental matter. This is enhanced to $E_n$ in the SCFT \cite{Seiberg:1996bd,Morrison:1996xf,Douglas:1996xp} (see also \cite{Kim:2012gu,Bashkirov:2012re,Mitev:2014jza,Tachikawa:2015mha,Chang:2017cdx,Chang:2017mxc} for further evidences).\footnote{Here $E_n$ for $n=1,2,\dots,5$ denotes $SU(2),SU(2)\times U(1),SU(3)\times SU(2),SU(5),SO(10)$ global symmetries respectively.}
	
	The 5d $\cN=1$ IR gauge theories have standard half-BPS boundary conditions preserving the 4d $\cN=1$ supersymmetry (see Appendix~\ref{app:5dsusy} for details). The hypermultiplet splits into two $\cN=1$ chiral multiplets on $\Sigma$, and setting either to zero leads to the boundary conditions $\cB_X[\Phi]$ and $\cB_Y[\Phi]$ defined in the last section. Similarly the 5d vector multiplet $V$ which contains a real scalar $\sigma$, a gauge field $A_\m$ and a gaugino $\lambda^i_\A$,
	decomposes into one 4d $\cN=1$ vector multiplet $v$ and one chiral multiplet $\phi$ of zero $U(1)_R$ charge. The supersymmetric Neumann and Dirichlet boundary conditions correspond to setting either $v$ or $\phi$ to zero 
	\ie
	\cB_N[V]:&\phi |_\Sigma =0 \to F_{ya}|_\Sigma=P_+\lambda^2|_\Sigma=\sigma+iA_y|_\Sigma=0\,,
	\\
	\cB_D[V]:& v |_\Sigma =0 \to A_a|_\Sigma=P_-\lambda^2|_\Sigma=D_y\sigma|_\Sigma=0\,.
	\label{VBNBD}
	\fe
	Here we study the maximally symmetric boundary conditions for the IR gauge theories of the $E_n$ SCFTs coming from assigning $\cB_D[V]$ or $\cB_N[V]$ to the 5d $SU(2)$ vector multiplet, and assigning $\cB_X[\Phi]$ to all $n-1$ fundamental hypermultiplets.\footnote{More general boundary conditions and interfaces for 5d $\cN=1$ gauge theories were considered in \cite{Gaiotto:2015una}, including a duality interface that maps one boundary to another while preserving the boundary 't Hooft anomalies. At the fixed point, such a duality interface should correspond to a superconformal interface in the 5d SCFT with vanishing $a$ and $c$ anomalies.
	} We expect them to be described by certain strongly coupled superconformal boundary conditions for the SCFT, which we will define as $\cB_N[E_n]$ and $\cB_D[E_n]$ respectively.\footnote{This is strongly supported by a nontrivial superconformal index on $S^1\times HS^4$ which counts boundary local operators in protected representations of the boundary superconformal symmetry $\mf{su}(2,2|1)$ \cite{Gaiotto:2015una}.} In the IR gauge theory, these boundary conditions preserve the $U(1)_I\times U(n-1) \subset U(1)_I \times SO(2n-2)$ subgroup of the bulk symmetry. It would be interesting to understand the symmetry enhancement in the presence of boundaries.  
	
	The boundary 't Hooft anomalies for $\cB_N[E_n]$ and $\cB_D[E_n]$ are matched by those of the gaugino $\lambda$ in the IR gauge theory.  Since $\lambda^2$ has $U(1)_R$ charge $-1$ and transforms in the adjoint representation of the $SU(2)$ gauge group, from Table~\ref{tab:bta}, we find 
	\ie
	\cI_6(\cB_N[V])=&{1\over 4} c_1(F)^3-{1\over 16} p_1(T) c_1(F)  +  c_1(F)c_2(F_G)\,,
	\\
	\cI_6(\cB_D[V])=&-{1\over 4} c_1(F)^3+{1\over 16} p_1(T) c_1(F) - c_1(F)c_2(F_G)\,.
	\label{Enba}
	\fe
	Note the mixed anomaly between $U(1)_R$ and the $G=SU(2)$ gauge symmetry.
	
	For the Dirichlet boundary condition $\cB_D[E_n]$, the bulk gauge symmetry $SU(2)$ becomes an emergent global symmetry on the boundary, but it cannot mix with the $U(1)_R$ due to its nonabelian nature. Consequently the 't Hooft anomalies for the superconformal $U(1)_R$ can be read off from \eqref{Enba},
	\ie
	k_{RRR}=k_R=-{3\over 2}\,,
	\fe
	and the boundary conformal anomalies follow from \eqref{ackrel},
	\ie
	a(\cB_D[E_n])=-{9\over 32},\quad c(\cB_D[E_n])=-{3\over 16}\,.
	\fe

	In the case of the Neumann boundary condition $\cB_N[E_n]$, since the $SU(2)$ gauge fields are dynamical on the boundary $\Sigma$, a mixed $U(1)_R$-$SU(2)$ anomaly would break the $U(1)_R$ symmetry explicitly. We can remedy this by introducing local degrees of freedom on the boundary. For example, we can couple the bare Neumann boundary condition $\cB_N[E_n]$ to $2N_f$ 4d chiral multiplets $Q_I$ that transform as doublets (with indices $I=1,2$) under the $SU(2)$ gauge group, and denote the modified boundary condition as $\cB_N^{N_f}[E_n]$.\footnote{The number of boundary fundamental chiral multiplets is chosen to be even to avoid the global Witten anomaly \cite{Witten:1982fp}. Relatedly the $SU(2)$ gauge theory in 5d has a discrete theta angle $\theta=0,1$ due to $\pi_4(SU(2))=\mZ_2$. Here this theta angle is  trivial $\theta=0$ for the $E_n$ theories. If $\theta=1$, there is a nontrivial inflow of the Witten anomaly to the boundary which must be cancelled for a Neumann type boundary condition (e.g. by introducing one more fundamental chiral multiplet on the boundary).}  These chiral multiplets provide an additional $U(2N_f)$ flavor symmetry from locally conserved currents on $\Sigma$
	and the $U(1)$ factor can mix with the $U(1)_R$ symmetry of the boundary. The superconformal $U(1)_R$ symmetry of $\cB_N^{N_f}[E_n]$ is the unique combination that is free from a $U(1)_R$-$SU(2)$ anomaly (see Table~\ref{tab:bta}), which requires assigning the following R-charge to the chiral multiplets, 
	\ie
	R(Q_I)=1-{1\over N_f }\,.
	\label{RQEn}
	\fe
	Correspondingly the $U(1)_R$ 't Hooft anomalies are
	\ie
	k_{RRR}={3\over 2}-{2\over  N_f ^2}\,,\quad 
	k_R={3\over 2}-2\,,
	\fe
	and the defect conformal anomalies are
	\ie
	a(\cB_N^{N_f}[E_n])={15\over 32}-{9\over 16_f^2}\,,\quad c(\cB_N^{N_f}[E_n])={1\over 2}-{9\over 16N_f^2}\,.
	\label{acBNEn}
	\fe
	Note that in the above we have assumed the absence of accidental $U(1)$ symmetries that can also mix with the $U(1)_R$ symmetry. One way to detect such phenomena is to check whether unitarity bounds are obeyed by operators with the putative R-symmetry \eqref{RQEn} \cite{Seiberg:1994pq,Intriligator:1995au}. Here the meson operator $M=\ep^{IJ}Q_I Q_J$ is a gauge invariant scalar chiral primary operator on the boundary whose conformal dimension is fixed by its R-charge,
	\ie
	\Delta(M)={3\over 2}R(M)=3 R(Q)=3-{3\over N_f}\,.
	\fe
	This is consistent with the 4d unitarity boundary $\Delta\geq 1$ for $N_f\geq 2$, which is a necessary condition for our results \eqref{acBNEn} to be physically meaningful. 
	
	Instead of adding fundamental 4d chiral multiplets on the boundary, one can also introduce other matter (or more generally a 4d $\cN=1$ SCFT with $U(1)$ global symmetries) to cancel the mixed $U(1)_R$\,-$SU(2)$ anomaly. In the presence of multiple $U(1)$ symmetries free from this mixed anomaly, the boundary $a$-maximization procedure will be needed to pick out the superconformal $U(1)_R$ symmetry (see the next section for a simple example).
	We leave this exercise to the interested readers.

	\subsubsection{Boundary SQCD}
	The $\cN=1$ supersymmetric QCD  (SQCD) is described by an $\cN=1$ $SU(N)$ super-Yang-Mills theory coupled to $N_f$ pairs of chiral multiplets $(Q_I,\tilde Q^I)$ with $I=1,2,\dots,N$ transforming in the fundamental and anti-fundamental representations of $SU(N)$. When the number of flavors lie in the conformal window ${3\over 2}N\leq N_f\leq 3N$, the SQCD is expected to flow to a 4d $\cN=1$ SCFT \cite{Seiberg:1994pq,Intriligator:1995au}. Here we shall describe a boundary analog of the SQCD theories where the dynamical gauge field propagates in a 5d bulk, which provides candidates of superconformal boundary conditions for the bulk SCFT in the UV. 
	
	The relevant bulk theory is described by a 5d $\cN=1$ $SU(N)_\kappa$ gauge theory with $N\geq 3$ and  Chern-Simons level $\kappa$. For $0\leq \kappa \leq N$, the UV completion is expected to be a 5d SCFT $\cT_{N,\kappa}$ with $U(1)_I$ global symmetry and the IR gauge theory arises from a symmetric mass deformation coupled to the $U(1)_I$ current multiplet \cite{Intriligator:1997pq}.\footnote{For special Chern-Simons level $\kappa=N$, the SCFT $\cT_{N,\kappa}$ develops an enhanced $SU(2)$ flavor symmetry from instanton operators charged under $U(1)_I$ \cite{Bergman:2013aca,Tachikawa:2015mha}.} The $U(1)_I$ symmetry is realized by the instanton current in the IR. 
	Let us consider a supersymmetric boundary condition for $\cT_{N,\kappa}$ by assigning Neumann boundary condition $\cB_N[V]$ (see \eqref{VBNBD}) to the 5d $SU(N)$ vector multiplet (which emerge in the IR gauge theory description). The 5d gaugino contributes the following boundary anomalies (from Table~\ref{tab:bta})
	\ie
	\cI_6(\cB_N[V])=&{N^2-1\over 12} c_1(F)^3-{N^2-1\over 48} p_1(T) c_1(F)  + {N\over 2} c_1(F)c_2(F_{SU(N)})
	+{\kappa\over 6} c_3(F_{SU(N)})
	\,.
	\label{QCDbta}
	\fe
	Since the gauge fields are dynamical on the boundary $\Sigma$, we need to add additional 4d matter to cancel the $SU(N)$ gauge anomalies as well as the mixed $U(1)_R$-$SU(N)$ anomalies. One way to achieve this is to couple $\cB_N[V]$ to $N_f$ chiral multiplets $Q_I$ and $N_f+\kappa$ chiral multiplets $\tilde Q^I$ transforming in the fundamental and anti-fundamental representations of $SU(N)$ respectively.
	Note that a novelty compared to the 4d SQCD is unequal number of ``quarks" and ``anti-quarks'' here, where the offset is due to the anomaly inflow from the 5d $SU(N)_\kappa$ Chern-Simons coupling.
	We refer to this boundary field theory as \textit{boundary SQCD} and assume that it descends from a superconformal boundary condition $\cB_N^{N_f}[\cT_{N,k}]$ for the 5d SCFT $\cT_{N,k}$ upon the supersymmetric $U(1)_I$ mass deformation. Below we will study the boundary conformal anomalies for $\cB_N^{N_f}[\cT_{N,k}]$. 
	
	The boundary matter has $U(N_f)\times U(N_f+\kappa)$ global symmetry. In particular, it contains $U(1)_A$ axial and $U(1)_B$ baryon symmetries, familiar in the study of 4d SQCDs \cite{Seiberg:1994pq,Intriligator:1995au}. We will denote their generators by $R_A$ and $R_B$ respectively. The chiral multiplets have charges
	\ie
	R_A(Q)=R_A(\tilde Q)=1\,,\quad R_B(Q)=-R_B(\tilde Q)=1\,.
	\fe
	The $U(1)_R$ symmetry relevant for the superconformal boundary $\cB_N^{N_f}[\cT_{N,k}]$, is generally a combination with parameters $t_A$ and $t_B$,
	\ie
	R_t= R_{5d} + t_A R_A +t_B R_B
	\fe
	where $R_{5d}$ is the R-symmetry inherited from the 5d bulk under which $Q$ and $\tilde Q$ are uncharged. 
	
	In order for the corresponding R-current to be conserved in the presence of dynamical $SU(N)$ gauge fields, we demand a vanishing mixed $U(1)_R$-$SU(N)$ anomaly,
	\ie
	{N\over 2}+ {N_f\over 2} (-1+ t_A +t_B)+{N_f+\kappa\over 2}(-1-t_A+t_B)=0\,.
	\label{QCDaf}
	\fe
	The 't Hooft anomalies for the candidate $U(1)_R$-symmetry follow from \eqref{QCDbta} and the boundary matter content,
	\ie
	k_{R_tR_tR_t}=&{N^2-1\over 2}+N_f(-1+ t_A +t_B)^3+(N_f+\kappa)(-1-t_A+t_B)^3\,,
	\\
	k_{R_t}=&{N^2-1\over 2}+N_f(-1+ t_A +t_B)+(N_f+\kappa)(-1-t_A+t_B)={N^2-2N-1\over 2}\,.
	\fe
	Carrying out the boundary $a$-maximization subject to the constraint \eqref{QCDaf}, we find that the trial anomaly $a(t)$ is maximized at 
	\ie
	t_B=0,\quad t_A=1-{N\over \kappa+2N_f}\,,
	\fe
	and the boundary conformal anomalies are
	\ie
	a(\cB_N^{N_f}[\cT_{N,k}])=&{3(N^2+N-1)\over 32}
	-{9 N^3\over 32(\kappa+2N_f)^2}\,,
	\\
	c(\cB_N^{N_f}[\cT_{N,k}])=&{2N^2+5N-2\over 32}
	-{9 N^3\over 32(\kappa+2N_f)^2}\,.
	\fe
	Once again, unitarity bound on the boundary meson operator $M=Q_I \tilde Q^I$ requires
	\ie
	\Delta(M)=3-{3N\over \kappa+2N_f} \geq 1 \,,
	\fe
	thus we should choose $N_f$ such that
	\ie
	2N_f\geq {3N\over 2}-\kappa\,.
	\fe
	It would be interesting to understand the fate of the boundary SQCD beyond this range. We leave this to future investigation.

	\subsection{Codimension-two defects in 6d SCFTs}
	Let us now discuss $p=4$-dimensional superconformal defects in 6d SCFTs.
	In 6d $\cN=(1,0)$ SCFTs, they correspond to half-BPS codimension-two defects preserving $\cN=1$ superconformal symmetry. For 6d $\cN=(2,0)$ SCFTs, both half-BPS and quarter-BPS codimension-two defects are present, preserving 4d $\cN=2$ and $\cN=1$ superconformal symmetries respectively. They play important roles in the class S construction of $\cN=2$ SCFTs in four dimensions \cite{Gaiotto:2009we,Gaiotto:2009hg} as well as the $\cN=1$ generalizations \cite{Bah:2011vv,Bah:2012dg,Xie:2013gma}.
	
	Up to mixing with $U(1)$ symmetries localized on the defect volume $\Sigma$, 
	the $U(1)_R$ symmetry of the codimension-two defect is identified with the following combination of symmetry generators in the 6d $\cN=(1,0)$ superconformal algebra $\mf{osp}(6^*|2)$,
	\ie
	R={2\over 3}(2R_{6d}-M_\perp)\,,
	\label{6d4dR}
	\fe
	where $R_{6d}$ is the Cartan element of the 6d $SU(2)_R$ symmetry normalized to have integer eigenvalues, and $M_\perp$ is the rotation generator in the transverse plane with eigenvalues $\pm {1\over 2}$ when acting on spacetime spinors.

	\subsubsection{Codimension-two defects in free theories}
	In the free 6d SCFT described by a free $\cN=(1,0)$ hypermultiplet $\Phi$ with holomorphic scalars $(X,Y)$ of scaling dimension $\Delta =2$, a half-BPS superconformal codimension-two defect can be defined by a scale invariant singularity of the form\footnote{This is an obvious generalization of the construction for 3d $\cN=4$ hypermultiplet in \cite{Dimofte:2019zzj} to higher dimensions.}
	\ie
	X(x_a,w)\sim {\A_X\over w^2}\,,\quad  Y(x_a,w)\sim {\A_Y\over w^2}\,,
	\label{hyperpole}
	\fe
	where $w$ is the complex coordinate for the transverse directions to the defect. The singularity is clearly invariant under the $U(1)_R$ symmetry \eqref{6d4dR}.
	Similar defects can be defined in the free $\cN=(2,0)$ SCFT using the hypermultiplet within the $\cN=(2,0)$ tensor multiplet.  We note that the singularity \eqref{hyperpole} implies the existence of a dimension zero operator on the defect worldvolume $\Sigma$ that carries nontrivial spin under the transverse rotation. This is somewhat unconventional and indicates that the naive cluster decomposition fails on $\Sigma$ \cite{Lauria:2020emq}.
	
	More generally, codimension-two defects in  free theories can be classified by studying  boundary conditions for the conformally coupled free fields on $AdS_5\times S^1$ with metric
	\ie
	ds^2=R^2{du^2+ dz_a^2\over u^2}+R^2 d\theta^2\,,
	\fe 
	which is related to flat space by a Weyl transformation. For free scalar fields this analysis was done in \cite{Rodriguez-Gomez:2017kxf,Nishioka:2021uef} and the conformal $a$-anomalies for the Dirichlet and Neumann boundary conditions were computed using the heat kernel method.
	Nontrivial superconformal codimension-two defects in the free 6d SCFTs correspond to supersymmetric completions of these boundary conditions on $AdS_5\times S^1$.
	In these cases, the conformal anomalies follow from the 't Hooft anomalies as in \eqref{ackrel}, which can be determined by inflow from the Kaluza-Klein tower of fermions and two-forms (from the 6d tensor multiplet) upon reduction on $S^1$. This setup can also be extended to interacting 6d SCFTs (see for example \cite{Aharony:2015zea}). We leave the study of such supersymmetric boundary conditions on $AdS_5\times S^1$ to future work.

	\subsubsection{Punctures in interacting SCFTs}

	More interesting defects arise in interacting 6d SCFTs. Despite the lack of  perturbative Lagrangians for such theories, the existence of various defects can be inferred by numerous constructions in string/M/F-theory, and by compactifying the 6d theory on compact manifolds and reducing to lower dimensional theories where a Lagrangian can become available. 
	The most well-studied examples are half-BPS codimension-two defects in the 6d $(2,0)$ SCFTs labelled by an ADE Lie algebra $\mf{g}$ \cite{Gaiotto:2009we,Gaiotto:2009hg,Chacaltana:2012zy}. The defects are characterized by homomorphisms $\varphi:\mf{su}(2) \to \mf{g}$, and so we will refer to them as $\cD_\varphi[\mf{g}]$. When the 6d SCFT is compactified on a Riemann surface $\cC$ with suitable twisting to preserve an $\mf{su}(2,2|2)$ subalgebra (which contains \eqref{defectalg}). These codimension-two defects can be added without further breaking the symmetry. They introduce punctures on the Riemann surface $\cC$ and contribute intimately to various aspects of the resulting 4d theory. In particular, the codimension-two defects are crucial to determining the 't Hooft and conformal anomalies of the 4d SCFTs (see \cite{Chacaltana:2012zy} for an extensive review). However a proper characterization of the conformal anomalies for defects was missing in these works, and the relations between the defect conformal and 't Hooft anomalies \eqref{ackrel} were assumed. Furthermore the defect 't Hooft anomalies were mostly inferred from consistency checks within the class S construction, and a direct derivation for the defect 't Hooft anomalies was not available until recently \cite{Bah:2018gwc,Bah:2018jrv,Bah:2019rgq,Bah:2019jts}.

	From the discussions in the previous sections, we now understand precisely what such defect anomalies mean in terms of the DCFT data (e.g. in \eqref{dta}). They are physically different from the anomalies of standalone CFTs. For example the classes of anomalies are much richer and conventional unitarity constraints on the anomalies no longer hold ($a$ and $c$ can be negative in unitary DCFTs).\footnote{The codimension-two defect also hosts nontrivial extrinsic conformal anomalies in addition to the $a$- and $c$-anomalies that depend on the extrinsic curvature. The conventional class S setup involves a direct product geometry $\cM_6=\cM_4\times \cC$ for the 6d theory and consequently such extrinsic anomalies do not contribute. They will be important if we were to generalize the class S setup by including a nontrivial warp factor.
	} Yet the defect anomalies still share many features that we are familiar with in the case of standalone CFTs, such as a monotonicity $a$-theorem which we have proved in Section~\ref{sec:dathm}. Furthermore we have also established firmly the anomaly multiplet relation \eqref{ackrel} and the $a$-maximization principle (see Theorem~\ref{thm:damax}) for these selected defect anomalies with superconformal symmetry. 
	
	In the following we will simply collect the recent results for defect 't Hooft anomalies from \cite{Bah:2019jts}, and restate the results, which follow from \eqref{ackrel}, as the defect conformal anomalies defined in \eqref{dta}.
	
	We will focus on the case $\mf{g}=A_{N-1}$ for which the work of \cite{Bah:2019jts} applies. Here $\rho$ is equivalent to a partition $[n_i]$ of $N$ with $N=n_1+\dots +n_k$ and $n_i\geq n_{i+1}>0$. The defect $\cD_{[n_i]}[A_{N-1}]$ can be engineered by a single M5 brane intersecting $N$ parallel M5 branes in a particular coincident limit. Alternatively, the same defect is described by $N$ M5 branes probing a Taub-NUT space ${\rm TN}_k$ with $k$-centers that collide in a singular limit \cite{Gaiotto:2009we,Gaiotto:2009hg,Tachikawa:2011dz}. Upon compactifying the 6d $(2,0)$ SCFT on $T^2$ which gives rise to the 4d $\cN=4$ super-Yang-Mills theory, this defect becomes a Gukov-Witten surface operator which has explicit Lagrangian descriptions \cite{Gukov:2006jk}.
	
	The authors of \cite{Bah:2019jts} determined the defect 't Hooft anomalies of $\cD_{[n_i]}[A_{N-1}]$ from inflow in M-theory using the second description of the defect above. The results were given in a different parametrization of the anomaly polynomial $\cI_6$ with
	\ie
	k_{RRR}={2\over 27}(n_v-n_h)+{8\over 9}n_v\,,\quad k_R={2\over 3}(n_v-n_h)\,,
	\fe
	and for the defect $\cD_{[n_i]}[A_{N-1}]$,\footnote{This comes from taking the difference between the ``inflow'' contribution and the ``non-puncture'' contribution from equations (6.2) in \cite{Bah:2019jts} and simplifying as explained therein.}
	\ie
	&(n_v-n_h)([n_i])={1\over 2}\left (N-\sum_{i=1}^{n_1} s_i^2 \right),
	\\
	&n_v([n_i])=  {1\over 6}N(N+1)(4N-1)-\sum_{i=1}^{n_1}\left(N^2- \left(\sum_{j=1}^i s_i\right)^2 \right)\,.
	\label{nvnhSUN}
	\fe
	Here $[s_i]$ with $1\leq i \leq n_1$ is the dual (transpose) partition of $[n_i]$. The defect $a$- and $c$-anomalies follow from \eqref{ackrel}, and coincide with their expected contributions to the 4d $\cN=2$ SCFT in the class S construction \cite{Chacaltana:2012zy}.\footnote{Note that with the enhanced $\cN=2$ superconformal symmetry on the defect, $a$-maximization is trivial.}
	
	The formula \eqref{nvnhSUN} has a natural generalization for general half-BPS codimension-two defects of the type $\cD_\varphi[\mf{g}]$ given in \cite{Chacaltana:2012zy}, leading to the following expressions for the defect conformal anomalies,
	\ie
	a(\cD_\varphi[\mf{g}])=& 2\rho_{\mf{g}}\cdot \rho_{\mf{g}}-\rho_{\mf{g}}\cdot h +{5\over 48} {\rm dim\,}\mf{g}_{1} 
	+{1\over 48} ({\rm rank\,}\mf{g}-{\rm dim\,} \mf{g}_0)\,,
	\\
	c(\cD_\varphi[\mf{g}])=& 2\rho_{\mf{g}}\cdot \rho_{\mf{g}}-\rho_{\mf{g}}\cdot h+{1\over 12} {\rm dim\,}\mf{g}_{1} 
	+{1\over 24} ({\rm rank\,}\mf{g}-{\rm dim\,} \mf{g}_0)\,.
	\label{gen20da}
	\fe
	Here $\rho_\mf{g}$ is the Weyl vector for $\mf{g}$, $h = \varphi(\sigma_3)$, and $\mf{g}$ is decomposed with respect to the eigenvalues of $[h,\cdot]$ as 
	\ie
	\mf{g}=\bigoplus_{j\in \mZ} \mf{g}_j\,.
	\fe
	To prove the formulas \eqref{gen20da} for codimension-two defects in general $(2,0)$ SCFTs requires a derivation of the corresponding defect 't Hooft anomalies, by extending the work of \cite{Bah:2019jts} to cases with an M-theory orientifold (for $\mf{g}=D_n$), and by studying inflow in IIB string theory with ADE singularities \cite{Bah:2020jas}.

	Before ending this section, we note that beyond the family of the $\cD_\varphi[\mf{g}]$ defects 
	which define regular (tame) punctures in the class S setup, the 6d $(2,0)$ SCFTs admit a much larger zoo of superconformal codimension-two defects that give rise to irregular (wild) punctures where the superconformal symmetry is emergent in the IR 
	\cite{Gaiotto:2009hg,Bonelli:2011aa,Xie:2012hs,Gaiotto:2012sf,Wang:2015mra,Xie:2017aqx,Nishinaka:2019nuy}, as well as the twisted defects (punctures) which are attached to codimension-one topological defects generating the outer-automorphism symmetry of certain $(2,0)$ theories \cite{Tachikawa:2010vg,Chacaltana:2012ch,Chacaltana:2012zy,Chacaltana:2014nya,Chacaltana:2015bna,Tachikawa:2018rgw,Wang:2018gvb,Beem:2020pry}.
	More recently, codimension-two defects in 6d $\cN=(1,0)$ SCFTs including the E-string theory have also been analyzed \cite{Heckman:2016xdl,Razamat:2016dpl,Kim:2017toz,Hassler:2017arf,Kim:2018bpg,Kim:2018lfo}. The results about their contributions to the conformal anomalies of the 4d SCFT in a generalized class S setup should again be interpreted as defect conformal anomalies in the sense explained here. 
	
	In complementary to the rich landscape of examples we have for codimension-two defects in 6d SCFTs, it would be interesting to understand and identify universal bounds on their physical data, much like what we have done in the case of  standalone 4d SCFTs, using the conformal bootstrap approach (see \cite{Poland:2018epd} for a review). For example, one may wonder if there is notion of \textit{minimal defect} that minimizes certain 't Hooft or conformal anomalies in a given bulk SCFT. Since such defects can be used to engineer 4d SCFTs upon compactification, this information will also be relevant for the search of minimal 4d SCFTs that have been explored in \cite{Poland:2015mta,Xie:2016hny,Buican:2016hnq,Li:2017ddj,Maruyoshi:2018nod}.
	
	\section{Discussions}
	\label{sec:discussion}
	In this paper, we have analyzed the anomalies of conformal defects (or DCFTs) of dimension $p=4$ in $d$-dimensional CFTs. We proved a defect analog of the 4d $a$-theorem which states that the defect conformal $a$-anomaly must decrease along unitary defect RG flows connecting UV and IR DCFTs. For conformal defects that preserve the minimal amount of supersymmetry, we established the anomaly multiplet relations between defect conformal $a$- and $c$-anomalies, and the 't Hooft anomalies involving the superconformal $U(1)_R$ symmetry. The general 't Hooft anomalies are determined by inflow from the bulk CFT, and the $U(1)_R$ symmetry is identified by the defect $a$-maximization principle which we have also derived. Together they provide a non-perturbative pathway to the conformal anomalies of strongly coupled defects. To illustrate our methods, we examined a number of examples of defects in 5d and 6d SCFTs. Here we conclude by discussing a few future directions beyond those mentioned in the main text.

	\subsubsection*{Defect correlation functions and defect chiral algebras}
	Conformal symmetry places stringent constraints on the correlation functions of local operators. In conventional CFTs in dimension $d\geq 4$, the two- and three-point functions of the stress-tensor $T_{ab}$ is completely fixed by conformal symmetry and Ward identities, up to three constants \cite{Osborn:1993cr},
	\ie
	\la T_{ab}(z) T_{cd}(0) \ra =&  c \la\la T_{ab}(z) T_{cd}(0) \ra\ra\,,
	\\
	\la T_{ab}(z_1) T_{cd}(z_2)T_{ef}(0) \ra 
	= & c \la\la T_{ab}(z_1) T_{cd}(z_2)T_{ef}(0) \ra\ra^c
	+
	a \la\la T_{ab}(z_1) T_{cd}(z_2)T_{ef}(0) \ra\ra^a
	\\
	&
	+
	b \la\la T_{ab}(z_1) T_{cd}(z_2)T_{ef}(0) \ra\ra^b\,.
	\fe
	Here $\la\la \cdot \ra\ra$ denotes theory independent conformal structures. For $d=4$, the coefficients $a$ and $c$ are nothing but the conformal anomalies defined in \eqref{trgen} for a 4d CFT. 
	
	In the presence of a $p$-dimensional conformal defect $\cD$, the correlators of the bulk stress-tensor $T_{\m\n}$ are constrained by the $SO(p,2)$ conformal symmetry and the $d$-dimensional Ward identities. Following the logic of \cite{Osborn:1993cr}, we expect for $p=4$, the defect $c$-anomaly to be determined by the defect two-point function $\la T_{\m\n}(x_1) T_{\rho\sigma}(x_2) \ra_\cD$, and the defect $a$-anomaly by the three-point function $\la T_{\m\n}(x_1) T_{\rho\sigma}(x_2)T_{\lambda\zeta}(x_3) \ra_\cD$.
	However because of the extra transverse directions, there are now additional conformally invariant tensor structures and furthermore their coefficients are general functions of invariant cross-ratios. The structure of the defect two-point function $\la T_{\m\n}(x_1) T_{\rho\sigma}(x_2) \ra_\cD$ has been worked in \cite{McAvity:1993ue,McAvity:1995zd,Herzog:2017xha,Herzog:2020bqw}. If $\cD$ is a conformal boundary (i.e. $d=p+1$), 
	this is determined by a single function $f(\xi)$ of the invariant cross-ratio $\xi$,
	\ie
	\xi\equiv {(x_1-x_2)^2\over y_1 y_2}\,.
	\fe
	For $p=4$, the conformal $c$-anomaly should be determined by (a limit of) $f(\xi)$ but the explicit relation is still to be derived.\footnote{One such relation was proposed in \cite{Herzog:2017kkj} but a counter-example appeared in \cite{Herzog:2020lel}.}
	For defects of higher codimensions, there is one more independent cross-ratio
	\ie
	\xi'\equiv {y_1\cdot y_2 \over |y_1| |y_2|}\,,
	\fe
	and the number of independent tensor structures is two for $d=p+2$ and seven for $d-p>2$ \cite{Herzog:2020bqw}. The defect three-point function  $\la T_{\m\n}(x_1) T_{\rho\sigma}(x_2)T_{\lambda\zeta}(x_3) \ra_\cD$ is much more complicated, with six cross-ratios in general and many tensor structures \cite{Guha:2018snh}.
	
	Supersymmetry are known to produce new (differential) constraints on these tensor structures. It would be interesting to explore the structure of stress-tensor multiplet correlation functions for superconformal defects. Furthermore, when sufficient supersymmetry is preserved, it is possible to define a simpler but nontrivial subsector of the full operator algebra in the DCFT that is closed under OPE. In particular, in the case of a half-BPS codimension-two superconformal defect in the 6d $(2,0)$ SCFT, the chiral algebra defined in \cite{Beem:2013sza} has a natural extension to defect operators with respect to the $\mf{su}(2,2|2)$ defect superconformal symmetry. The resulting chiral algebras will be of a different nature. For example the absence of a local stress-tensor multiplet on the defect implies that the corresponding chiral algebra no longer contains a Virasoro subalgebra. Furthermore the bulk operators in the 6d $(2,0)$ SCFT also contain a chiral algebra subsector defined with respect to a different $D(2,2)$ subalgebra of the 6d superconformal algebra $\mf{osp}(8^*|4)$ \cite{Beem:2014kka}. The interplay between these protected subsectors of bulk and defect operators will provide a wealth of information in the $\cN=2$ supersymmetric DCFTs. We hope to report on this in the future.

	\subsubsection*{Bounds on defect conformal anomalies}
	As we have emphasized in the main text, despite sharing many features of the conformal anomalies of standalone CFTs, the defect conformal anomalies are physically distinct. In particular, in a unitary DCFT, both the $a$- and $c$-anomalies can be negative, and no lower bounds have been identified. In light of these observations, perhaps we should look for bounds on the ratio $a\over c$ of the defect conformal anomalies. 
	
	For conventional CFTs, such bounds arise naturally in studying positivity constraints of the energy correlators in a normalized state created by local operators, and are known as the conformal collider bounds \cite{Hofman:2008ar} (see also \cite{Hofman:2016awc}). For general unitary CFTs, the conformal anomalies are constrained by
	\ie
	{31\over 18}\geq {a\over c}\geq {1\over 3}\,.
	\fe
	If the CFT is superconformal, a stronger bound is achieved, depending on the amount of supersymmetry preserved,
	\ie
	{\cN=1}:~{3\over 2}\geq {a\over c}\geq {1\over 2}
	\,,\quad 
	{\cN=2}:~{5\over 4}\geq {a\over c}\geq {1\over 2}\,.
	\fe
	In all cases, the upper and lower bounds are saturated by free vector and scalar theories respectively, with appropriate supersymmetric completions.
	It would be very interesting to explore a generalization of the collider bounds in \cite{Hofman:2008ar} to cases with conformal boundaries or more general conformal defects, by studying positivity constraints on energy correlators in the presence of defect excitations. We emphasize that the simplest superconformal boundary condition for a 5d hypermultiplet has a vanishing defect $c$-anomaly (see Section~\ref{sec:5dbdyfree}). Therefore in order for such bounds to exist in $d=5$, extra restrictions on the defect (boundary) need to be imposed. We have not observed similar issues in $d=6$.

	\section*{Acknowledgements}
	The author thanks Nathan Agmon  for collaborations on related topics. The author is also grateful to Zohar Komargodski, Ken Intriligator and Yuji Tachikawa for reading and commenting on a draft of the manuscript. 
	The work of YW is  supported in part by the Center for Mathematical Sciences and Applications and the Center for the Fundamental Laws of Nature at Harvard University.

	\appendix
	
	\section{Boundary four-point amplitude in free scalar theory}
	\label{app:integral}
	Here we study the large $m$ expansion of the one-loop Feynman diagram that computes the four-point function of $\Phi^2(z,0)$ on the Neumann boundary of a free scalar field in $d=5$,
	\ie
	I_{1234}=&\int {d^4k \over (2\pi)^4}
	{1\over (|k|+m) (|k+p_1|+m)(|k+p_1+p_2|+m)( |k-p_4|+m)}\,.
	\fe
	We start by introducing Schwinger parameters for the propagators,
	\ie
	I_{1234}=&
	\int {d^4k \over (2\pi)^4} \int \prod_{i=1}^4 ds_ie^{-m\sum_i s_i}
	e^{-(s_1 |k|+s_2|k+p_1|+s_3|k+p_1+p_2|+s_4|k-p_4|)}\,.
	\fe
	Next we use the Laplace transform
	\ie
	\int_0^\infty {dt\over 2\sqrt{\pi}} {s e^{-{s^2\over 4 t}}e^{-tk^2}\over t^{3/2}}=e^{-s|k|}\,,
	\fe
	and obtain
	\ie
	I_{1234}=&
	\int {d^4k \over (2\pi)^4} \int \prod_{i=1}^4 ds_ie^{-m\sum_i s_i}\int \prod_{i=1}^4 {dt_i\over 2\sqrt{\pi}} 
	e^{-(t_1  k^2+t_2|k+p_1|^2+t_3|k+p_1+p_2|^2+t_4|k-p_4|^2)}\prod_{i=1}^4{s_i e^{-{s_i^2\over 4 t_i}}\over t_i^{3/2}}\,.
	\fe
	Performing the $k$ integral, this gives 
	\ie
	I_{1234}=& 
	{1\over 2^8 \pi^{6}} \int \prod_{i=1}^4 ds_ie^{-m\sum_i s_i}\int \prod_{i=1}^4 dt_i 
	\prod_{i=1}^4{s_i e^{-{s_i^2\over 4 t_i}}\over t_i^{3/2}}
	{\pi^2\over  (\sum_i t_i)^2}e^{t_1 t_3 s+ t_2 t_4 t\over\sum_i t_i  }\,,
	\label{Isim}
	\fe
	after Wick rotating to Minkowski signature and imposing the ``on-shell'' condition $p_i^2=0$.
	
	Let us now expand $I_{1234}$ in the large $m$ limit. We are particularly interested in the four-derivative term which takes the following form as is clear from the symmetry of \eqref{Isim},
	\ie
	I_{1234}\supset {\A_1(s^2+t^2)+ \A_2 s t \over m^4}\,.
	\fe
	Expanding the last exponential factor in \eqref{Isim}, we obtain
	\ie
	\A_1=& 
	{1\over 2(4\pi)^4}  \int \prod_{i=1}^4 ds_ie^{-\sum_i s_i}\int \prod_{i=1}^4 dt_i 
	\prod_{i=1}^4{s_i e^{-{s_i^2\over 4 t_i}}\over t_i^{3/2}}
	{(t_1 t_3)^2\over  (\sum_i t_i)^4} \,.
	\fe
	Performing a change of variables $s_i \to t_i s_i$, this becomes 
	\ie
	\A_1
	=&
	{1\over 2(4\pi)^4 }      \int_0^\infty  \prod_{i=1}^4 ds_i s_i\int_0^\infty \prod_{i=1}^4 dt_i 
	\prod_{i=1}^4{  e^{-{t_i\over 4}({s_i^2}+4s_i)} t_i^{1\over 2}}
	{(t_1 t_3)^2\over  (\sum_i t_i)^4} \,.
	\label{alpha1}
	\fe
	The $t_i$ integral can be simplified using the following integration identity, from {\bf 4.638} in \cite{ID} (we have corrected a typo there),
	\ie
	\int_0^\infty \prod_{i=1}^n dt_i e^{-\sum_i q_i t_i}{\prod_i t_i^{p_i-1}\over (\sum_i t_i)^r}
	={\prod_i\Gamma (p_i)\over \Gamma(r)}\int_0^\infty dx  {x^{r-1}\over \prod_i (x+ q_i)^{p_i}},
	\label{ID}
	\fe
	with $q_i,p_i,r>0$ and $\sum_i p_i>r$.
	
	Applying \eqref{ID} to \eqref{alpha1}, we find
	\ie
	\A_1=&
	{1\over 2(4\pi)^4 }    \int_0^\infty  \prod_{i=1}^4 ds_i s_i\int_0^\infty dx {\Gamma\left(7\over 2\right)^2\Gamma\left(3\over 2\right)^2\over \Gamma(4)}{x^3\over \prod_{i=1}^2 ({1\over 4}s_i^2+s_i+x)^{7\over 2}\prod_{i=3}^4 ({1\over 4}s_i^2+s_i+x)^{3\over 2}}\,.
	\fe
	This last integral can be evaluated in \texttt{Mathematica} by first integrating $s_i$ and then $x$, giving
	\ie
	\A_1 =&\frac{61}{645120 \pi ^2}\,.
	\fe
	A similar computation also determines $\A_2$,
	\ie 
	\A_2 =&\frac{1}{107520 \pi ^2}\,.
	\fe
	Combining with the contributions from the other two one-loop diagrams $I_{1342}$ and $I_{1423}$, we find that the full four-point amplitude at the fourth derivative order is given by
	\ie
	I_{1234}+I_{1342}+I_{1423}\supset \frac{17}{92160 \pi ^2} {s^2+t^2 +u^2 \over m^4}\,.
	\label{4pfinal}
	\fe

	\section{Supersymmetric boundaries for 5d $\cN=1$ gauge theories}
	\label{app:5dsusy}
	
	\subsection{5d Spinor conventions}
	The 5d Gamma matrices $\CC_\m$ satisfy the Clifford algebra
	\ie
	(\CC_\m)^\A{}_\C(\CC_\n)^\C{}_\B+(\CC_\n)^\A{}_\C(\CC_\m)^\C{}_\B=2\eta_{\m\n}\D_\B^\A\,,
	\fe
	where $\A,\B=1,2,3,4$ are the 5d Dirac spinor indices. 
	Their transpose $\CC_\m^t$ and conjugate $\CC_\m^*$ obey the same algebra and are related to $\CC_\m$ by
	\ie
	\CC_\m^t=C\C_\m C^{-1}\,,\quad -\CC_\m^*=B\CC_\m B^{-1}\,.
	\fe
	Here $C$ and $B$ are charge conjugation matrices related by $C=B\CC^0$ and satisfy
	\ie
	C_{\A\B} = -C_{\B\A}\,,\quad (C^{-1})^{\A\B}=-(C^*)^{\A\B}\equiv C^{\A\B}\,.
	\fe
	The symplectic-Majorana (SM) condition on a 5d spinor reads
	\ie
	(\Psi_A^\A)^*=\ep^{AB} C_{\A\B} \Psi_B^\B\,,
	\fe
	with the convention $\ep^{12}=\ep_{21}=1$.
	
	\subsection{Hypermultiplet}
	A 5d $\cN=1$ hypermultiplet consists of four real scalars $\Phi^{iA}$, a SM fermion $\Psi^A$ and four auxiliary real scalars $F^{iA}$. Here $i=1,2$ and $A=1,2$ are the $SU(2)_R$ and $SU(2)_F$ doublet indices respectively. These indices are lowered and raised by the invariant tensors $\ep_{ij},\ep^{ij},\ep_{AB},\ep^{AB}$ satisfying $\ep^{ij}\ep_{jk}=\D^i_k$ and $\ep^{AB}\ep_{BC}=\D^A_C$.
	The fields are subject to the reality conditions
	\ie
	(\Phi^{iA})^*=\ep_{AB}\ep_{IJ} \Phi^{jB}\,,\quad 
	(\Psi^{A}_\A)^*=\ep_{AB}C^{\A\B} \Psi^{B}_\B\,,\quad 
	(F^{iA})^*=\ep_{AB}\ep_{IJ} F^{jB}\,.
	\fe
	The on-shell supersymmetry transformations are
	\ie
	\D_\xi \Phi^{iA}=-2i \xi^i \Psi^A,\quad \D_\xi \Psi^A= \CC^\m \xi_i \pa_\m \Phi^{iA}\,,
	\label{hypersusy}
	\fe
	where $\xi^i_\A$ is a SM spinor corresponding to the eight supercharges. 
	
	We identity the 4d $\cN=1$ superalgebra by the following projection 
	\ie
	{\bm\xi}= P_+ \xi^1\,,
	\label{4dn1susy}
	\fe
	where $P_\pm \equiv {1\over 2}(1\pm\CC_y)$. Then it's clear from \eqref{hypersusy} that the hypermultiplet splits into two chiral multiplets closed under $\D_{\bm \xi}$ separately,
	\ie
	X=(\Phi^{11},P_+\Psi^1,\pa_y \Phi^{21}),\quad 
	Y=(\Phi^{12},P_+\Psi^2,\pa_y \Phi^{22})\,.
	\fe
	Note that $\pa_y \Phi^{21}$ and $\pa_y \Phi^{22}$ coincide with the on-shell auxiliary field $F^{21}$ and $F^{22}$ respectively. This comes from the effective 4d superpotential $\int dy \int d^2\theta X\pa_y Y $ from the 5d Lagrangian of a hypermultiplet on $\mR^{4,1}_+$ \cite{Dimofte:2012pd}.
	
	The consistent boundary conditions preserving boundary $\cN=1$ supersymmetry amounts to setting a linear combination of $X$ and $Y$ to zero identically. They define the $\cB_X[\Phi]$ and $\cB_Y[\Phi]$ boundary conditions (and their rotated versions) in Section~\ref{sec:5dbdyfree}.
	
	\subsection{Vector multiplet}
	A 5d $\cN=1$ vector multiplet $V$ contains a real scalar $\sigma$, a SM fermion $\lambda_\A^i$, a gauge field $A_\m$ and three auxiliary scalars $D_{ij}$ with $i,j$ indices symmetrized and satisfying
	\ie
	(D_{ij})^* =D^{ij}=\ep^{ik}\ep^{jl}D_{kl}\,.
	\fe
	The supercharges act on the vector multiplet fields as
	\ie
	\D_\xi A_\m =&i \xi_i \C_\m \lambda^i\,,
	\\
	\D_\xi \sigma =& i  \xi_i \lambda^i\,,
	\\
	\D_\xi \lambda^i=&-{1\over 2}\CC_{\m\n}\xi^i F_{\m\n} +\CC^\m \xi^i D_\m \sigma +\xi_i D^{ij}\,,
	\\
	\D_\xi D^{ij}=&-2i \xi^{(i} \sD \lambda^{j)} + 2[\sigma, \xi^{(i} \lambda^{j)}]\,.
	\fe
	Under the 4d $\cN=1$ subalgebra generated by $\D_{\bm \xi}$ from \eqref{4dn1susy}, the vector multiplet splits into a chiral multiplet and a vector multiplet
	\ie
	v=(A_a,P_-\lambda^2, D_{12})
	,\quad
	\phi=( \sigma+i A_5, P_+\lambda^2,D_{11},D_{22})\,.
	\fe
	The consistent boundary conditions preserving boundary $\cN=1$ supersymmetry correspond to setting either $v$ or $\phi$ to zero, which defines the supersymmetric Dirichlet and Neumann boundary conditions of the 5d $\cN=1$ vector multiplet $V$,
	\ie
	\cB_N[V]:&\quad F_{ay}=\sigma+iA_y=P_+\lambda^2=0\,,
	\\
	\cB_D[V]:&\quad A_{a}=D_y\phi=P_-\lambda^2=0\,.
	\fe
	\bibliographystyle{JHEP}
	\bibliography{defREF,SYMdefect}

\end{document}